\documentclass{svjour3}                     
\smartqed  
\usepackage{graphicx,epsfig,amssymb}
\usepackage{natbib}
\newcommand{\ion}[2]{#1\,{\sc{#2}}}
\newcommand{\calG}{$\mathcal{G}$}
\newcommand{\calR}{$\mathcal{R}$}
\newcommand{\arcsec}{\mbox{$^{\prime\prime}$}}
\journalname{SSRv}
\begin{document}

\title{He-like ions as practical astrophysical plasma diagnostics:
}
\subtitle{From stellar coronae to active galactic nuclei}

\author{D.\ Porquet         \and
        J.\ Dubau  \and
        N.\ Grosso 
}
\authorrunning{Porquet et al.} 

\institute{Delphine Porquet and Nicolas Grosso \at
              Observatoire Astronomique de Strasbourg, Universit{\'e} de
Strasbourg, CNRS, UMR 7550,\\ 11 rue de l'Universit{\'e}, F-67000
Strasbourg, France\\
              \email{delphine.porquet@astro.unistra.fr, nicolas.grosso@astro.unistra.fr}           
           \and
           Jacques Dubau \at
             Institut d'Astrophysique Spatiale, Universit{\'e} Paris
Sud-11, F-91405 Orsay Cedex, France\\
              \email{jacques.dubau@ias.u-psud.fr}
              }

\date{Received: date / Accepted: date}

\maketitle

\begin{abstract}
We review X-ray plasma diagnostics based on the line ratios of
He-like ions. Triplet/singlet line intensities can be used
to determine electronic temperature and density, and were  first
developed for the study of the solar corona. Since the launches of the
X-ray satellites Chandra and
XMM-Newton, these diagnostics have been extended  
and used (from \ion{C}{v} to \ion{Si}{xiii}) for a wide variety of
astrophysical plasmas such as stellar coronae, supernova remnants, solar
system objects, active galactic nuclei, and X-ray binaries. Moreover, the
intensities of He-like ions can be used to determine the ionization
process(es) at work, as well as the distance between the X-ray plasma
and the UV emission source for example in hot stars. 
In the near future thanks to the next generation of X-ray satellites
(e.g., Astro-H and IXO), higher-Z He-like lines (e.g., iron) will be
resolved, allowing plasmas with higher temperatures and densities to be probed. 
Moreover, the so-called satellite lines that are formed closed to
parent He-like
lines, will provide additional valuable diagnostics to determine 
electronic temperature, ionic fraction, departure from ionization
equilibrium and/or from Maxwellian electron distribution.  
\end{abstract}

\tableofcontents

\section{Introduction}\label{intro}

\begin{figure}[!t]
\begin{center}
  \includegraphics[height=0.5\columnwidth]{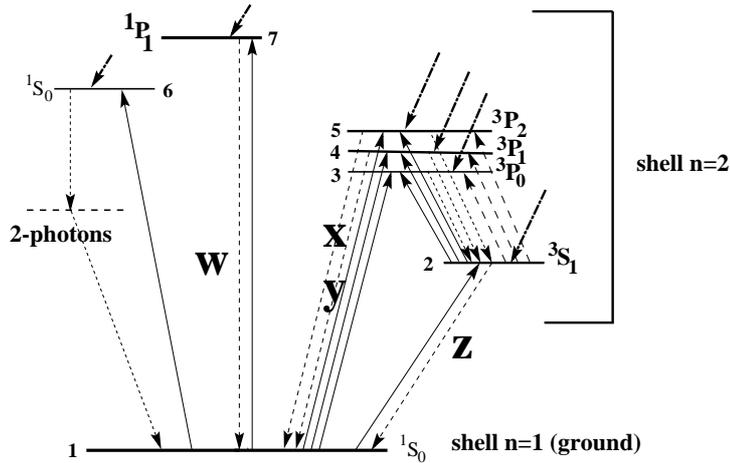}
\end{center}
\caption{Simplified level scheme for He-like ions. $w$ (or $R$),
$x,y$ (or $I$), and $z$ (or $F$) are the resonance, intercombination, and
forbidden lines, respectively. {\it Upward arrows} correspond to the 
electron collisional excitation transitions (solid) and to the
 photo-excitation from 2\,$^{3}$S$_{1}$ to 2\,$^{3}$P$_{0,1,2}$ levels (dashed). 
{\it Downward arrows} correspond to the 
radiative transitions (including 2-photon continuum from
2\,$^{1}$S$_{0}$ to the ground level). 
{\it Thick downward arrows}:
recombination (radiative and dielectronic) plus cascade
processes. Figure from \cite{PD00}. Courtesy of Astronomy \&
Astrophysics.}
\label{fig:grotrian}
\end{figure}

Spectral lines of H-like and He-like ions are among the most prominent 
features in X-ray spectra from a large variety of astrophysical
sources. Compared to other ionic iso-electronic sequences, 
He-like ions are abundant over the widest temperature range 
in collisional plasmas due to their closed shell ground state. 
The most intense He-like lines correspond to transitions between the
$n$ = 2 shell and the $n$ = 1 ground state shell (see
Fig.~\ref{fig:grotrian}):
\begin{itemize}
\item  The resonance line, named in the literature either $R$
or $w$ is an electric dipole transition (E1; 1s$^2$ $^{1}$S$_{0}$ --
1s\,2p $^{1}$P$_{1}$);
\item   The intercombination line $I$ is composed of two lines $x$ (M2:
magnetic quadrupole transition; 1s$^{2}$ $^{1}$S$_{0}$ -1s\,2p $^{3}$P$_{2}$) and $y$
(E1; 1s$^{2}$ $^{1}$S$_{0}$ - 1s\,2p $^{3}$P$_{1}$).   
The quadrupole line $x$ only becomes intense for He-like ions 
heavier than \ion{S}{xv}, and with the same intensity as the line $y$
for \ion{Ca}{xix}. The transition from $^{3}$P$_{0}$ cannot decay
to the ground level since this transition is strictly forbidden but
decays to the $^{3}$S$_{1}$ metastable level.
\item The forbidden line $F$ or $z$ is a relativistic
magnetic dipole transition (M1; 1s$^{2}$ $^{1}$S$_{0}$ - 1s\,2s $^{3}$S$_{1}$). 
\end{itemize}

\noindent The 1s\,2s $^{1}$S$_{0}$ level decays to the ground level by a
two-photon process (see section~\ref{sec:atomic}). The energies and
wavelengths of these lines for \ion{C}{v} (Z=6), \ion{N}{vi}, \ion{O}{vii},
\ion{Ne}{ix}, \ion{Mg}{xi}, \ion{Si}{xiii}, \ion{S}{xv},
\ion{Ca}{xix}, and \ion{Fe}{xxv} (Z=26) are given in Table 1. \\

These He-like lines were first observed in laboratory for C, F, Mg, Al
 \cite[see][]{Edlen47} and later in solar 
plasmas thanks to the Orbiting Solar Observatory ({\sl OSO}), and
rocket experiments 
\citep[e.g.,][]{Fritz67,Doschek70,Acton72,Grineva73}. However the detection of a
significant line coincident with the wavelength of a single photon
transition from the metastable $^{3}$S$_{1}$  level to the ground
level could not be understood from theory, because the
metastable level was expected to decay only by two-photon emission. 
\cite{Gabriel69a} strongly suggested that this line might correspond to
some unknown photon transition. This was confirmed by a
quantum-relativistic calculation of \cite{Griem69} who proved that
indeed the $^{3}$S$_{1}$ level  
has a significant single-photon decay rate, so a line can be
observed. Then \cite{Gabriel69b} proposed that the relative
intensities of these lines can be used for temperature and density
diagnostics (see section~\ref{sec:Tne}) for solar plasma:  \\ 
\begin{equation}\label{eq:R}
{\cal R}~(n_{\rm e})~\equiv~\frac{z}{x+y} ~~~~ \left({\rm or} \equiv~\frac{F}{I}\right)
\end{equation}
\begin{equation}\label{eq:G}
{\cal G}~(T_{\rm e})~\equiv~\frac{z+(x+y)}{w} ~~~~ \left({\rm or} \equiv~\frac{F+I}{R}\right)
\end{equation}
These diagnostics have first been widely used for solar spectra
 \citep[e.g.,][]{Doschek70,Acton72,McKenzie80,Wolfson83,Keenan84,Doyle86} 
and for X-ray spectra of tokamak plasmas \citep[e.g.,][]{Kallne83,Keenan89}. \\

\begin{table}[!t]
\begin{center}
\begin{tabular}{cccccccccc}
\hline
\hline
 & \ion{C}{v}&\ion{N}{vi}&\ion{O}{vii}  &\ion{Ne}{ix}
&\ion{Mg}{xi}  &\ion{Si}{xiii}&\ion{S}{xv}&\ion{Ca}{xix} &\ion{Fe}{xxv} \\
\hline
w         & {\tiny 40.2674}     & {\tiny 28.7870}     & {\tiny 21.6015}        &  {\tiny 13.4473}       & {\tiny 9.1688}        &  {\tiny 6.6479} & {\tiny 5.0387}     & {\tiny 3.1772} & {\tiny 1.8504}       \\
          & {\tiny (0.3079)}  & {\tiny (0.4307)}  & {\tiny
(0.5740)}     &  {\tiny (0.9220)}    & {\tiny (1.3522)}   & {\tiny (1.8650)}  & {\tiny (2.4606)} & {\tiny (3.9024)} & {\tiny (6.7004)}  \\  
x       & {\tiny 40.7280}     & {\tiny 29.0819}     & {\tiny 21.8010}        & {\tiny 13.5503}        &{\tiny 9.2282}         &{\tiny 6.6850} & {\tiny 5.0631}  & {\tiny 3.1891}  & {\tiny 1.8554}         \\
          & {\tiny (0.3044)}  & {\tiny (0.4263)}  & {\tiny (0.5687)}     & {\tiny (0.9150)}     & {\tiny (1.3431)}   &{\tiny (1.8547)}   & {\tiny (2.4488)} & {\tiny (3.8878)}& {\tiny (6.6823)} \\
y       & {\tiny 40.7302}     & {\tiny 29.0843}     & {\tiny 21.8036}        & {\tiny 13.5531}        &{\tiny 9.2312}         &{\tiny 6.6882}  & {\tiny 5.0665}    & {\tiny 3.1927}  & {\tiny 1.8595}      \\
          & {\tiny (0.3044)}  & {\tiny (0.4263)}  & {\tiny (0.5686)}     & {\tiny (0.9148)}     & {\tiny (1.3431)}   &{\tiny (1.8538)}   & {\tiny (2.4471)} & {\tiny (3.8833)}& {\tiny (6.6676)} \\
z         & {\tiny 41.4715}     & {\tiny 29.5347}     & {\tiny 22.0977}        & {\tiny 13.6990}        & {\tiny 9.3143}        & {\tiny 6.7403} & {\tiny 5.1015}  & {\tiny 3.2110}    & {\tiny 1.8682}     \\
          & {\tiny (0.2990)}  & {\tiny (0.4198)}  & {\tiny (0.5611)}     & {\tiny (0.9051)}     & {\tiny (1.3311)}   &{\tiny (1.8394)}  &  {\tiny (2.4303)}& {\tiny (3.8612)} & {\tiny (6.6366)} \\  
\hline
\hline
\end{tabular}
\end{center}
\vspace*{-0.2cm}
\caption{Wavelengths (in \AA), and in parentheses energies (in
keV) of the resonance ($w$), intercombination ($x+y$) et forbidden
($z$) lines for several He-like ions.}
\label{lambda}
\end{table}

It is now possible, thanks to the spectral resolution and the
sensitivity of the
 current generation of X-ray satellites {\sl Chandra} and {\sl XMM-Newton},
to resolve the He-like ion lines and to use these diagnostics for extra-solar objects.
 Indeed, the He-like ion line ratios are valuable tools
in the analysis of high-resolution spectra of a variety of plasmas such as: 
\begin{itemize}
\item Collisional Ionization Equilibrium (CIE) plasmas or also called coronal plasmas:\\
in such plasmas, ionization is due to electron-ion collisions and the
atomic levels are populated 
mainly by electron impact. It is commonly assumed that CIE plasmas are 
 optically thin to their own radiation, and that there is no external
radiation field that affects the ionization balance. However, in some
cases, these assumptions are not fulfilled as discussed in
sections~\ref{sec:processdiagn} and \ref{sec:Tne}.\\ 
{\it E.g., solar and stellar coronae (OB stars, late type stars, active stars,
T Tauri, ...)}, {\it cluster of galaxies}, {\it the hot
intra-cluster medium}, {\it Galactic
ridge} and {\it Galactic center X-ray emission, ...} \\

\item  Recombination-dominated or Photo-Ionization Equilibrium (PIE) plasmas:\\
in such plasmas, ionization is due to photons (ionizing radiation) and
the atomic levels are populated mainly by radiative recombination of
H-like ions to He-like ions directly or by cascade from upper levels. 
These plasma are generally overionized relative to
the local electronic temperature and have a much smaller electronic
temperature compared to CIE plasmas. That is why collisional excitations out of the
ground state are inefficient and excited levels are populated
via radiative recombination. However, as we will see in
section~\ref{sec:processdiagn}, photo-excitation can be a
non-negligible process. \\
{\it E.g., Active galactic nuclei, X-ray binaries, ...}\\
 
\item Out of equilibrium plasmas, or non-ionization equilibrium (NIE) plasmas:
Some astrophysical plasmas depart from ionization equilibrium . 
This occurs when one or several physical conditions of the plasma
suddenly change, such as the
temperature (section~\ref{sec:NIE}), the density, or the photo-ionization radiation field.\\
{\it E.g., Supernova remnants, 
solar and stellar flares, colliding winds in star clusters and X-ray
binaries, cluster of galaxies, intra-cluster medium in merging galaxy clusters, ...}.
\end{itemize}

Since the pioneering work of \cite{Gabriel69b}, several works have been
dedicated to the improvements of these diagnostics based on He-like
line ions and their 
extension to other types of plasmas
(photo-ionization equilibrium and non-ionization equilibrium): e.g.,
\cite{Mewe72}, \cite{Blumenthal72}, \cite{Gabriel73},
\cite{Mewe75,Mewe78a,Mewe78b,Mewe78c}, \cite{Acton78},
\cite{Pradhan81}, \cite{Bely-Dubau82}, \cite{Pradhan82,Pradhan85},
\cite{Keenan84}, \cite{Swartz93}, \cite{Liedahl99}, \cite{PD00,P01},
\cite{Bautista00}, \cite{Porter07}, and \cite{Smith09}.\\

In the present paper, we review why and how the relative intensities of the He-like 
ion lines can be used for plasma diagnostics. We also present several
observational results based on these diagnostics for different types of plasmas.
Since a large number of papers dealing with He-like ions has been published, this
review cannot be exhaustive.\\

The outline of this paper is as follows. First, we give a brief overview of 
atomic structure and a few basic processes that play an important role
for the population of the upper level of the $n$=2 shell in He-like
ions (sect.~\ref{sec:atomic}). 
Section~\ref{sec:processdiagn} concerns diagnostics of the
different ionization processes:  
 CIE and PIE plasmas (sect.~\ref{sec:CIEPIE}), NIE plasmas
(sect.~\ref{sec:NIE}), and charge-transfer (sect.~\ref{sec:CE}). 
Section~\ref{sec:Tne} is dedicated to electronic temperature
(sect.~\ref{sec:Te}) and electronic density 
(sect.~\ref{sec:ne}) diagnostics. Section~\ref{sec:satlines} presents the
possible diagnostics based on the satellite lines. 
In the last section, we conclude and bring some possible perspectives
for the future of plasma
diagnostics based on He-like ions and their satellite lines 
thanks to the next generation of
X-ray satellites, such as {\sl Astro-H} and {\sl IXO}.

\newpage
\section {Atomic processes and atomic data}\label{sec:atomic}

Here we briefly introduce the main atomic processes that lead to the formation
of He-like ion lines, as well as the impacts of atomic data, atomic 
model and spectral resolution on the calculation of the He-like line
ratios. For more details about X-ray spectroscopy and atomic
processes, see e.g., the very nice reviews from 
 \cite{Liedahl99}, \cite{Mewe99}, \cite{Paerels03}, \cite{Kahn05} and \cite{Kaastra08}.

\subsection{Main atomic processes}

As illustrated on the Grotrian diagram reported in Fig.~\ref{fig:grotrian}
(see also Fig.~1 in \citealt{Mewe78a}), the atomic levels of the He-like ions 
can be populated and depopulated by several atomic processes. \\ 

\noindent{\it Collisional excitation inside He-like ions}\\
In most plasmas, collisional excitations are mainly due to (projectile) 
electrons, however  collisional excitation by protons and
$\alpha$-particles can be important in some cases (see below).

Electron collisional excitations from the 1s$^2$ $^{1}S_{0}$ (ground) 
level to excited levels ($n$=2 and higher) require a large projectile
energy to open the 1s$^2$ shell. They 
become efficient as temperature increases and favors the population of
singlet levels, such as  $^{1}$P$_{1}$ level  (hence the resonance $w$
line). The excitation process of singlet excited levels from the
singlet  ground state does not require a change of the target spin. On
the contrary, excitation of  triplet levels is only possible by
exchange of the projectile electron with one of the target
electrons. As projectile energy increases, the exchange process becomes less
efficient than the direct process, i.e., the non-exchange process.   For
highly ionized He-like ions, the spin-orbit interaction becomes important
: spin-orbit interaction between singlet and triplet levels, for
example 1s\,2p $^{3}$P$_{1}$  mixed with 1s\,2p $^{1}$P$_{1}$. It is
responsible for the similar temperature behavior for $^{1}$P$_{1}$ and
$^{3}$P$_{1}$ excitations.  At high temperature, radiative cascade
contributions from $n >2$ levels populate the $n\ge$ 2 levels,  cascades
remaining inside singlet levels or triplet levels respectively. Due to
the small radiative  probabilities from the $n$=2 triplet level to the singlet
ground level, this favors the populations of triplet levels, namely
$^{3}$P$_{0,1,2}$ and $^{3}$S$_{1}$, compared to the singlet levels
which can decay more directly to the ground level. 
At low temperature, the contribution of the auto-ionizing resonances  
to the electron scattering cross-sections enhances the forbidden $z$ (and
in a smaller part the intercombination ones, $x$ and $y$)
 transition far more than the resonance transition \citep{Pradhan81}. 

Excitations inside the $n$ = 2 shell should be taken into
account even for low temperature plasmas, i.e., for photo-ionized plasmas.
First, the metastable $^{3}S_{1}$ level can be depopulated to the 
$^{3}P_{0,1,2}$ levels when the density is high enough, i.e., above
the so-called critical density (that depends on the He-like ions, see
section~\ref{sec:ne}). At much higher density, the 1s\,2s $^{1}$S$_{0}$ level
(upper level of the 2-photon transition) can be depopulated to 1s\,2p
$^{1}$P$_{1}$, thereby increasing the intensity of the resonance line
\citep{Gabriel72}. 

The calculations of proton impact excitation rates by \cite{Blaha71}
show that their contributions (and in a smaller part those of 
$\alpha$-particles) can be non-negligible for high-$Z$ (i.e., $>$14) 
ions at very high temperature \citep{Mewe78a}. However, new
calculations of these proton excitation rates are required (Dubau et
al., in preparation). \\

\noindent {\it Recombinations from H-like ions to He-like ions}\\ 
Recombinations from H-like ions to He-like ions can be due to
radiative recombination or dielectronic recombination. 
Radiative recombination is highly efficient at low temperature (few eV)
such as in photo-ionized plasmas, and favors the populations 
of the triplet levels, due to their higher statistical weight 
compared to singlet level. On the contrary, dielectronic recombination is 
efficient for high temperature plasmas, but it also favors the triplet levels. 
It is negligible in the low temperature range (i.e., photo-ionized plasmas). Hence, 
both recombination processes lead to an intense forbidden or intercombination lines 
(depending on the density, see section~\ref{sec:ne}) compared to the resonance line.\\

\noindent {\it Inner-shell ionization of Li-like ions}\\
Inner-shell ionization of Li-like ions can significantly
contribute to the formation of the forbidden line of He-like ions,
hence increasing the value of the \calR\ ratio at the low density limit  and the
value of the \calG\ ratio
\citep[e.g.,][]{Doschek70,Gabriel72,Mewe75,Mewe78a,Oelgoetz04}. 
For this process to have an impact on the intensity of the forbidden He-like ions, 
both the relative abundance of Li-like to He-like,
N(Li-like)/N(He-like), and the ionization coefficient rate must be large. 
This latter condition is reached at high temperature. 
In collisional ionization equilibrium plasmas and close to the temperature of 
maximum formation of He-like ions, both conditions are not fulfilled since the 
relative abundance of Li-like ions is very small.  
A high abundance of Li-like ions and a high electronic temperature
can occur in transient plasmas such as a in supernova remnants and
during solar/stellar flares (section~\ref{sec:NIE}), and this process
is important especially for high-$Z$ ions.  \\

\noindent{\it Other atomic and physical processes}\\
Other atomic processes should be considered in some cases such as
photo-excitation (section~\ref{sec:ne}) or charge exchange
(section~\ref{sec:CE}). 
For very high densities, not considered here, several atomic processes 
have to be taken into account \citep[e.g.,][]{Bautista00}.  
In addition, resonance line scattering and optical depth might have an
impact on the line ratios (section~\ref{sec:CIEPIE}). 

\subsection{Importance of the accuracy of the atomic model and atomic data}

As shown by several authors, to perform line
ratio calculations it is of importance to use a good atomic model with 
accurate atomic data \cite[e.g.,][]{Mewe78a,Pradhan81,PD00,P01,Bautista00,
Smith01,Smith09,Porter07}.  \\

It is not possible to use a $LS$ model of He-like states to simulate 
the intensities of the resonance, intercombination and forbidden lines even for 
low charge He-like ions. Nevertheless some fine-structure data can be converted from 
$LS$ data, particularly collisional data due to the non-relativistic nature 
of the main electron-electron interaction, $1/r_{ij}$. Be aware,
however it can not be used as a general rule because relativistic
effects increase rapidly 
as the nuclear charge increases. A second point concerns the number of He-like levels 
included in the model, and maybe H-like and Li-like levels as well. The $n=2$ triplet levels,
$^3$S$_1$, $^3$P$_{0,1,2}$ being very sensitive to recombination from
H-like ions, the model
must include high $n>$2 levels cascading to $n=2$ levels, even at low 
temperature. As temperature increases, radiative cascades contribution from collisional 
excitation of high $n>$2 levels can have also a significant influence on the populations 
of the $n=2$ levels \citep[e.g.,][]{PD00,Smith09}. Electron ionization
from Li-like ions 
is also possible and has a direct influence on the forbidden line intensity. Besides, 
proton and $\alpha$-particle excitations might play a role on the density diagnostic for
high-$Z$ He-like ions. A reliable simulation therefore requires collecting first a huge
amount of accurate atomic data related to $n=2$ and also to $n >2$, up
to $n=5$, or may be $n$=10. Papers giving all these data 
with great accuracy do not exist. Some papers contain apparently very accurate data for electron 
excitation but for only very few transitions. Further papers assert that the
preceding calculations 
are not complete because they do not contain some important effects, such as radiation 
damping of resonances, which invalidate their accuracy. On the other
hand, some authors  
give apparently less accurate data but including these effects. To illustrate this last point, 
we mention the impressive work of \cite{Sampson83}, for electron collisional excitation 
of He-like excited levels for $n=1$ and $n=2$ up to $n=5$, for $4\le
Z\le 74$. 
It is a Coulomb-Born-Exchange (CBE) calculation between 
fine-structure levels, apparently not very accurate. In particular, auto-ionizing resonances 
are not included. But they were inserted, including also radiation damping, in a following 
work by \cite{Zhang87}. 
Indeed, auto-ionizing resonances have to be taken into account for a good calculation of 
collisional excitation rates \citep[e.g.,][]{Pradhan81,Smith09}. But how to include 
them correctly ? There exists two different approaches either implicitly or explicitly. 
CBE, already mentioned, or Distorted Wave (DW) data do not include at all auto-ionizing 
resonances but it is possible to include them explicitly afterward. Whereas in 
Close-Coupling approximations, such as R-matrix (non-relativistic, Breit-Pauli or Dirac 
formalisms), they are implicitly included. These later approximations are therefore apparently better. 
But these resonance effects can be strong and, sometimes they are strongly
overestimated because auto-ionizing resonances can also decay by radiative transitions
not included in the approximations, the so-called radiation damping effect. The radiation
damping is also responsible for Dielectronic Recombination, included in the model but
as a recombination process (see a later section). One must therefore takes care not to 
include twice the same process, i.e., to separate the two decays of resonance either as
excitation or as recombination.
Many different methods  have been proposed to overcome this problem of radiation damping
in close-coupling calculations, \cite[e.g.,][]{Zhang95}. They can give
different results. What are the best ? 
The comparison between all these atomic data obtained by different
methods is very interesting 
but beyond the scope of this review. 

Experimental measurements of the atomic data and/or 
line ratio of He-like ions in laboratory devices, such as tokamak, EBIT
\citep[e.g.,][]{Kallne83,Keenan89,Beiersdorfer92,Wargelin08,Beiersdorfer09,Brown09}
could be of great importance to resolve some discrepancies between
theoretical calculations and observations
\citep[e.g.,][]{Ness03b,Testa04a,Chen06,Smith09}.  
As a summary on this point, the most important is first to have a good
atomic model containing the best data.

\subsection{Impact of the spectral resolution}

Blending of the He-like lines with satellite lines (defined in
section~\ref{sec:satlines}) depends on the resolution of the observed 
spectra. Therefore, the calculations of the line ratio have to take
into account contributions from unresolved satellite lines, especially
for high-$Z$ ions \cite[see e.g.,][]{P01,Sylwester08}. At low and moderate 
spectral resolutions all satellite lines are
unresolved. In the near future
 higher spectral resolution  will be obtained thanks to X-ray calorimeters and gratings
(aboard e.g., Astro-H and IXO) at high
energies and  some $n$=2-3 satellites lines will be resolved for
high-$Z$ ions, and can be used to probe plasma properties (see 
section~\ref{sec:satlines}). However $n>3$ satellite lines will not be
resolved and the \calG\ and \calR\ ratio calculations must account for
their contributions. Additionally, 
possible contamination could be due to other elements such as
\ion{Fe}{xix} lines with the intercombination line of \ion{Ne}{ix}
lines for high enough temperature \citep{McKenzie80,McKenzie85,Wolfson83, Ness03b}.

\section{Ionization process diagnostic}\label{sec:processdiagn}

\subsection{Collisional-dominated versus photo-ionized plasmas}\label{sec:CIEPIE}

\begin{figure}[!t]
\begin{center}
\begin{tabular}{cc}
\includegraphics[width=4cm,angle=0]{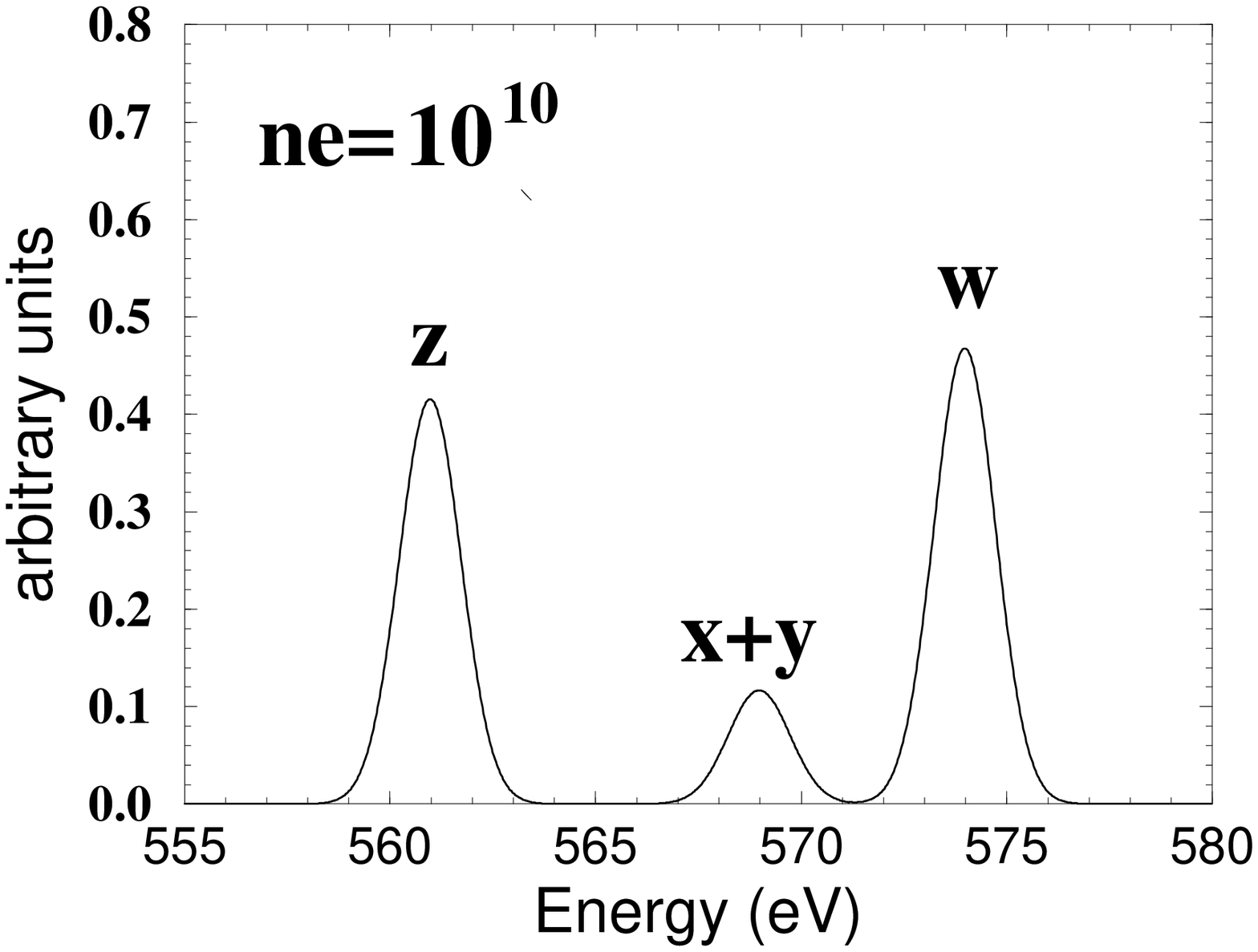} &\includegraphics[width=4cm,angle=0]{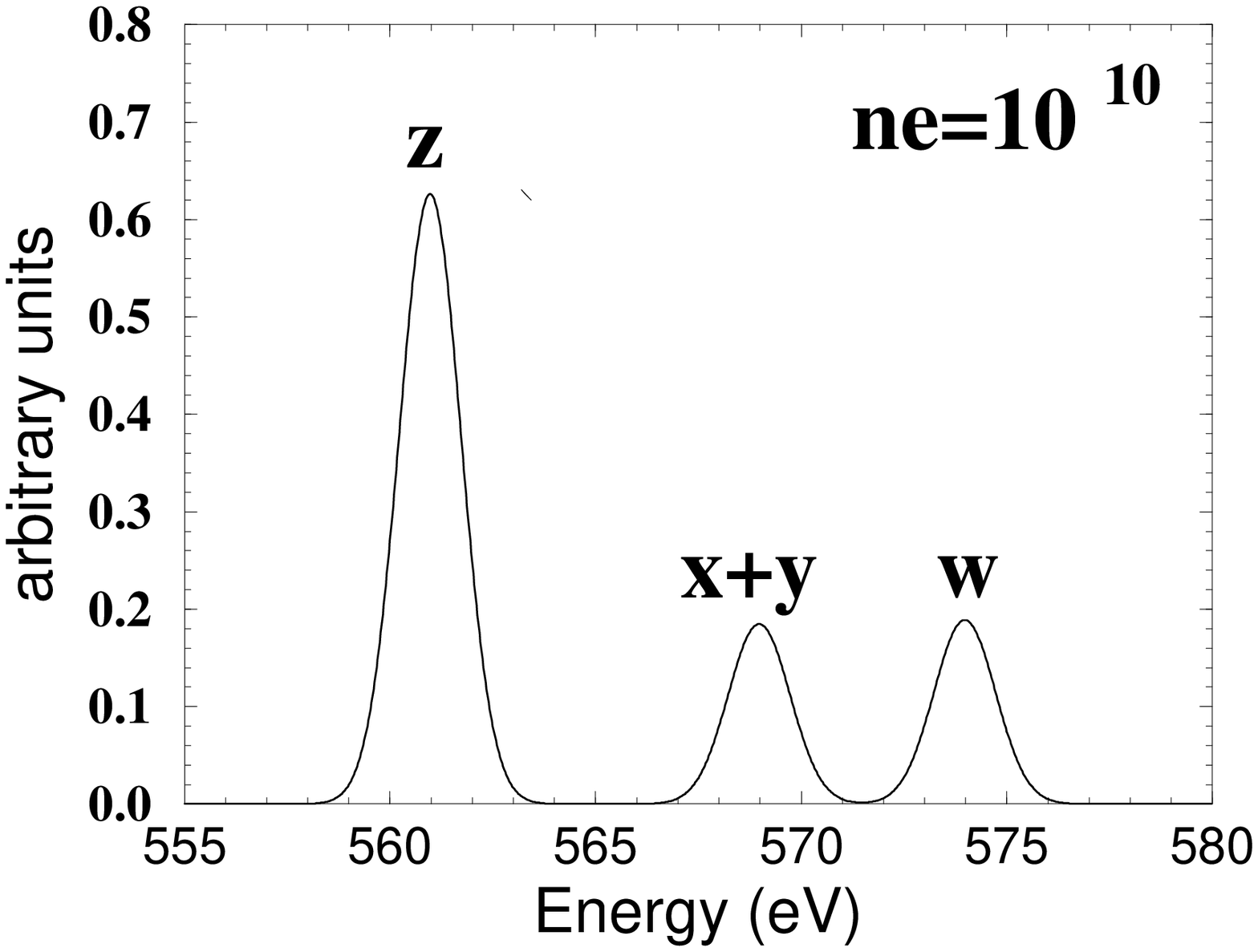}\\
\includegraphics[width=4cm,angle=0]{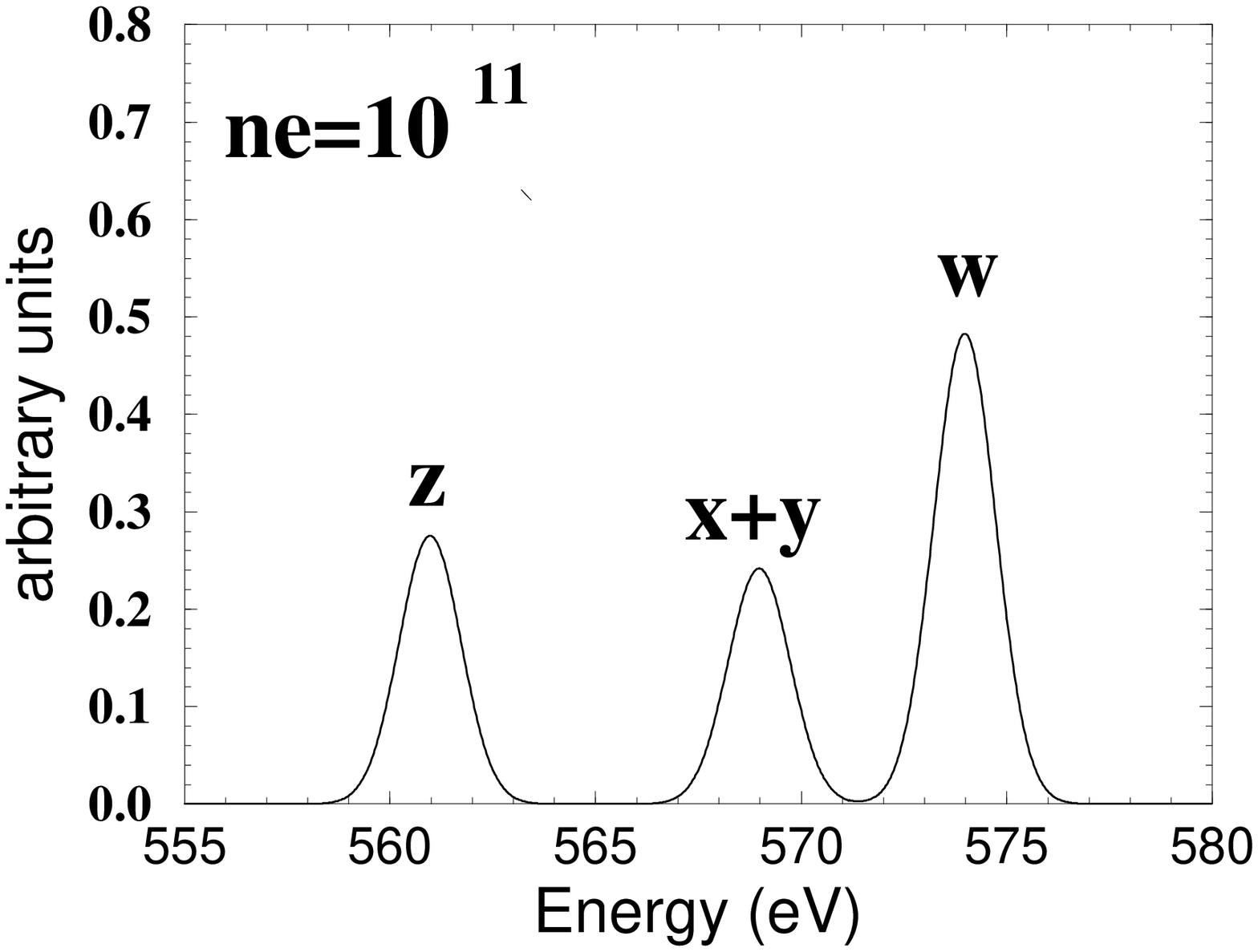} &\includegraphics[width=4cm,angle=0]{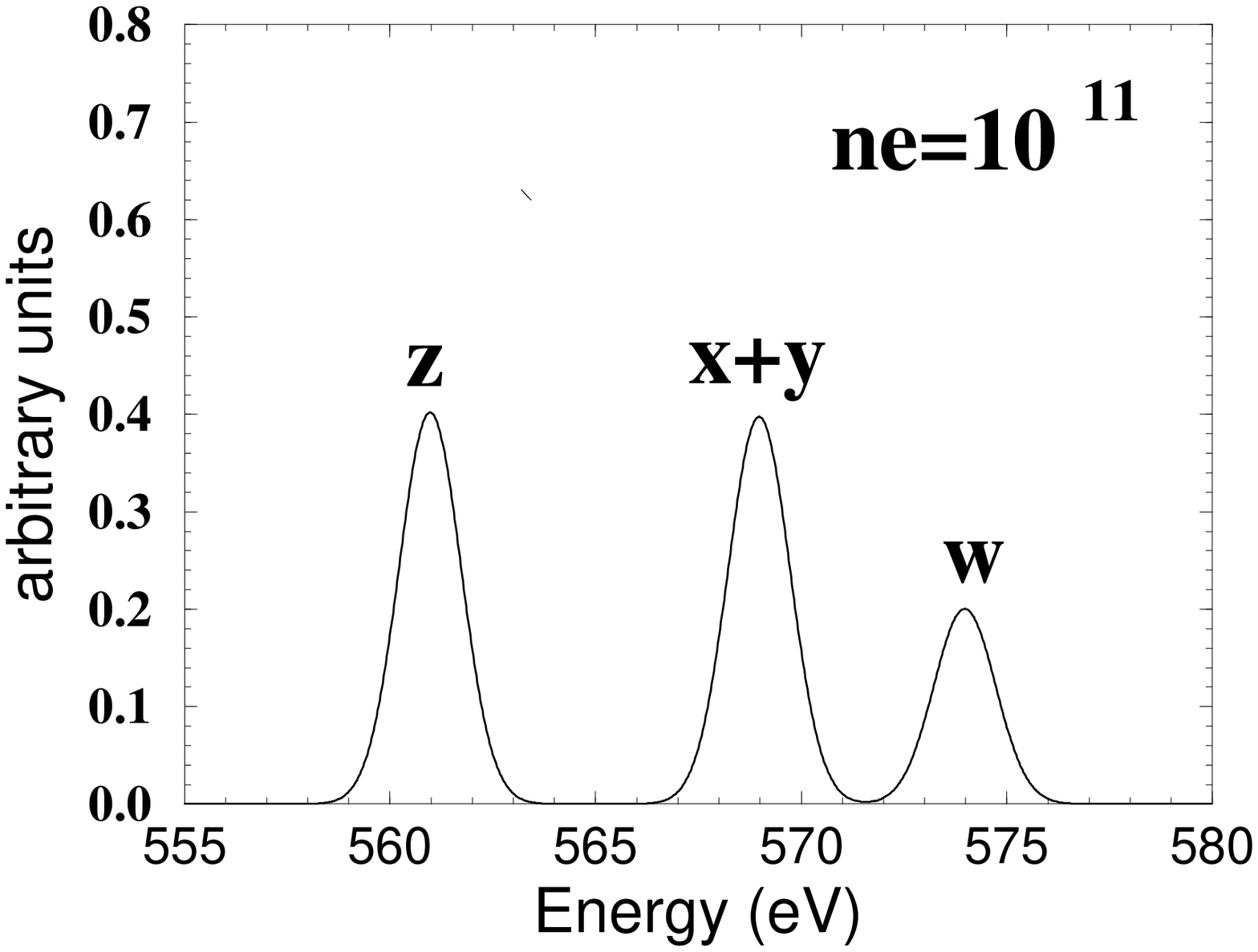}\\
\includegraphics[width=4cm,angle=0]{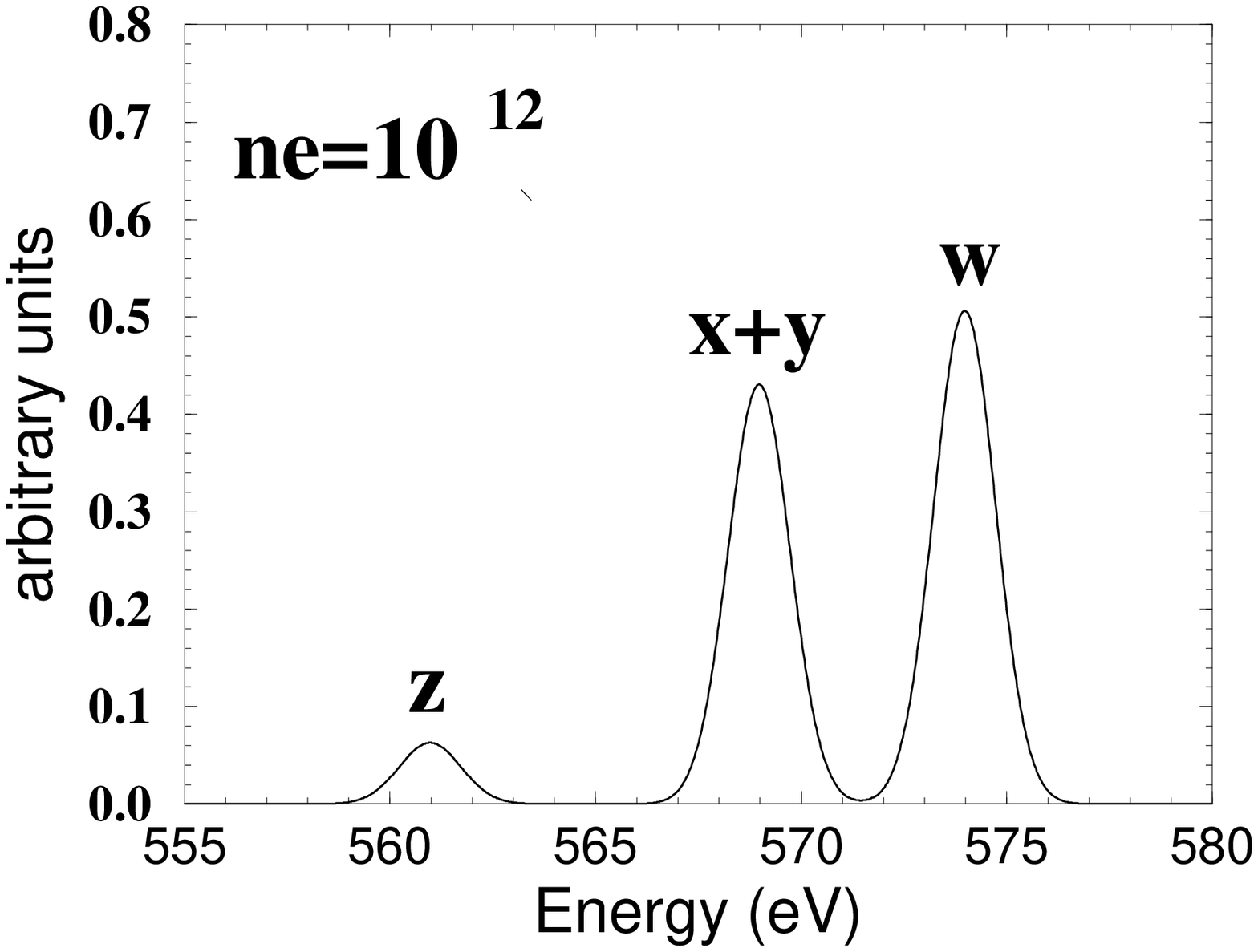} &\includegraphics[width=4cm,angle=0]{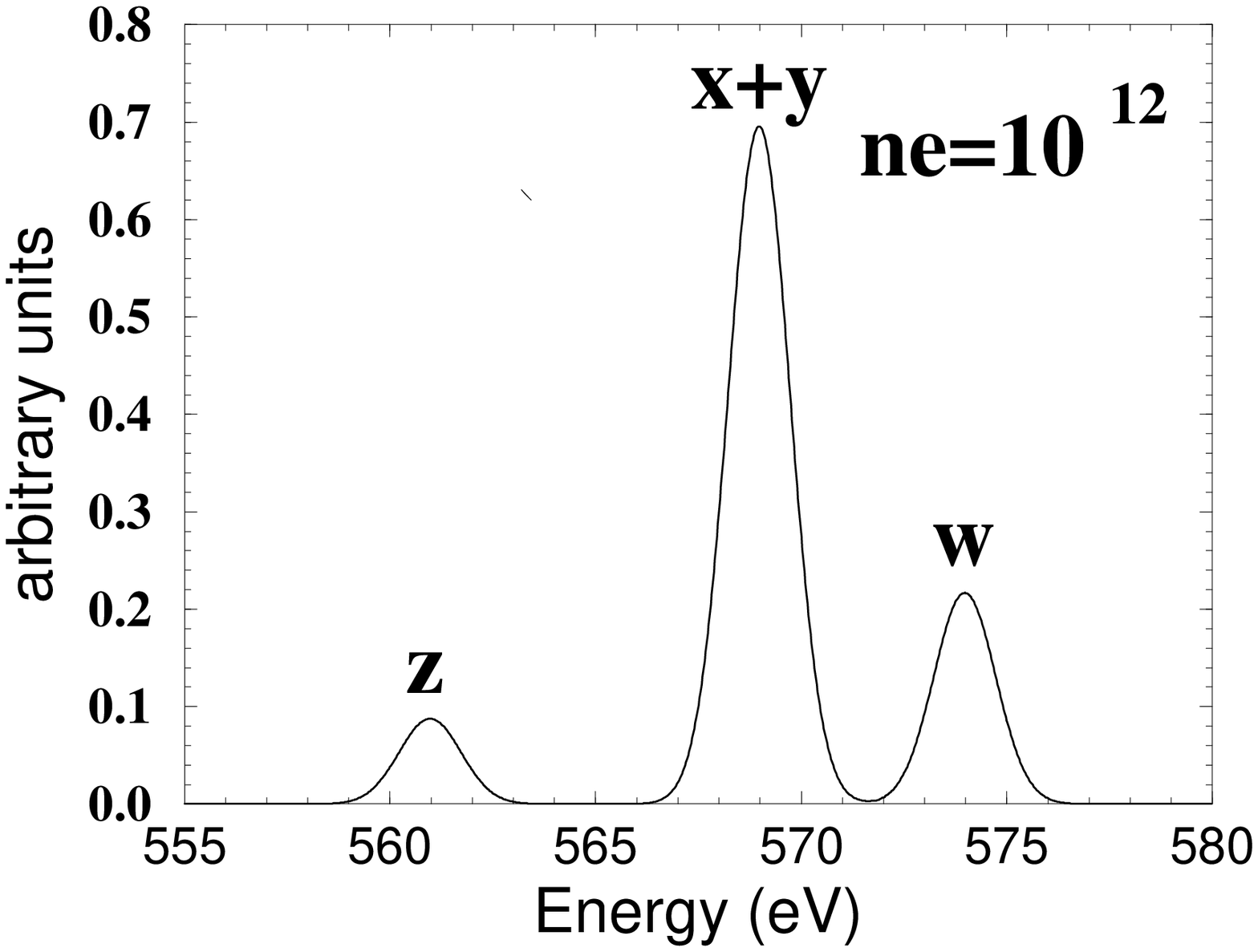}\\
\end{tabular}
\end{center}
\caption{Comparison of theoretical spectra of the \ion{O}{vii} triplet
lines at the spectral resolution of the RGS aboard {\sl XMM-Newton}. 
$z$, $x+y$ and $w$ mark the forbidden, intercombination and resonance
lines, respectively. Left and right panels show 
collisional-dominated 
 and photo-ionized (recombination-dominated)
plasmas, respectively. 
The three values of electronic density 
corresponds to the density range where the relative intensities of $z$ and
$x+y$ line vary. 
The intensities are normalized so as the sum of the lines to be equal to unity. 
Figure from \cite{PD00}. Courtesy of Astronomy and Astrophysics.}
\label{fig:OVIIlines}
\end{figure}

The relative intensity of the three lines can be used to discriminate
between a collisional-dominated plasma (CIE) and a photo-ionized 
(recombination-dominated) plasma (PIE). 
At high electronic temperature, as in collisional plasma, 
the collisional excitation process is efficient.   
Therefore this favors the population of the $^1P_1$ level, hence the
resonance line $w$ is intense. The value of the \calG\ ratio is around
1 (see section~\ref{sec:Te}). While at low electronic temperature, as
in photo-ionized plasma,  
 recombination is the dominant process. Recombination favors the population
of the $^3P_{0,1,2}$ and $^3S_1$ levels, hence the forbidden and/or
the intercombination lines are intense. The value of the \calG\ ratio
is equal or greater to 4. To illustrate this effect, the relative
intensities of the \ion{O}{vii} triplet lines for 
collisional and photo-ionized plasmas
are shown in Fig.~\ref{fig:OVIIlines}. 
This diagnostic has been used for active galactic nuclei (e.g., NGC\,4051:
\citealt{Collinge01}; NGC\,4593: \citealt{McKernan03}; NGC\,4151:
\citealt{Schurch04}) and X-ray binaries (e.g., EXO\,0748-67:
\citealt{Cottam01,Jimenez03}) where 
\calG\ greater than or similar to 4 were found, 
confirming the photo-ionization as the main
ionizing process in these objects.  
However this diagnostic based on the value of the \calG\ ratio 
to disentangle between a collisional plasma
and a photo-ionized plasma should be used with caution when 
 {\it resonance line scattering} and optical depth are not
negligible. 

\begin{figure}[!t]
\begin{center}
\begin{tabular}{cc}
\includegraphics[width=0.45\columnwidth,angle=0]{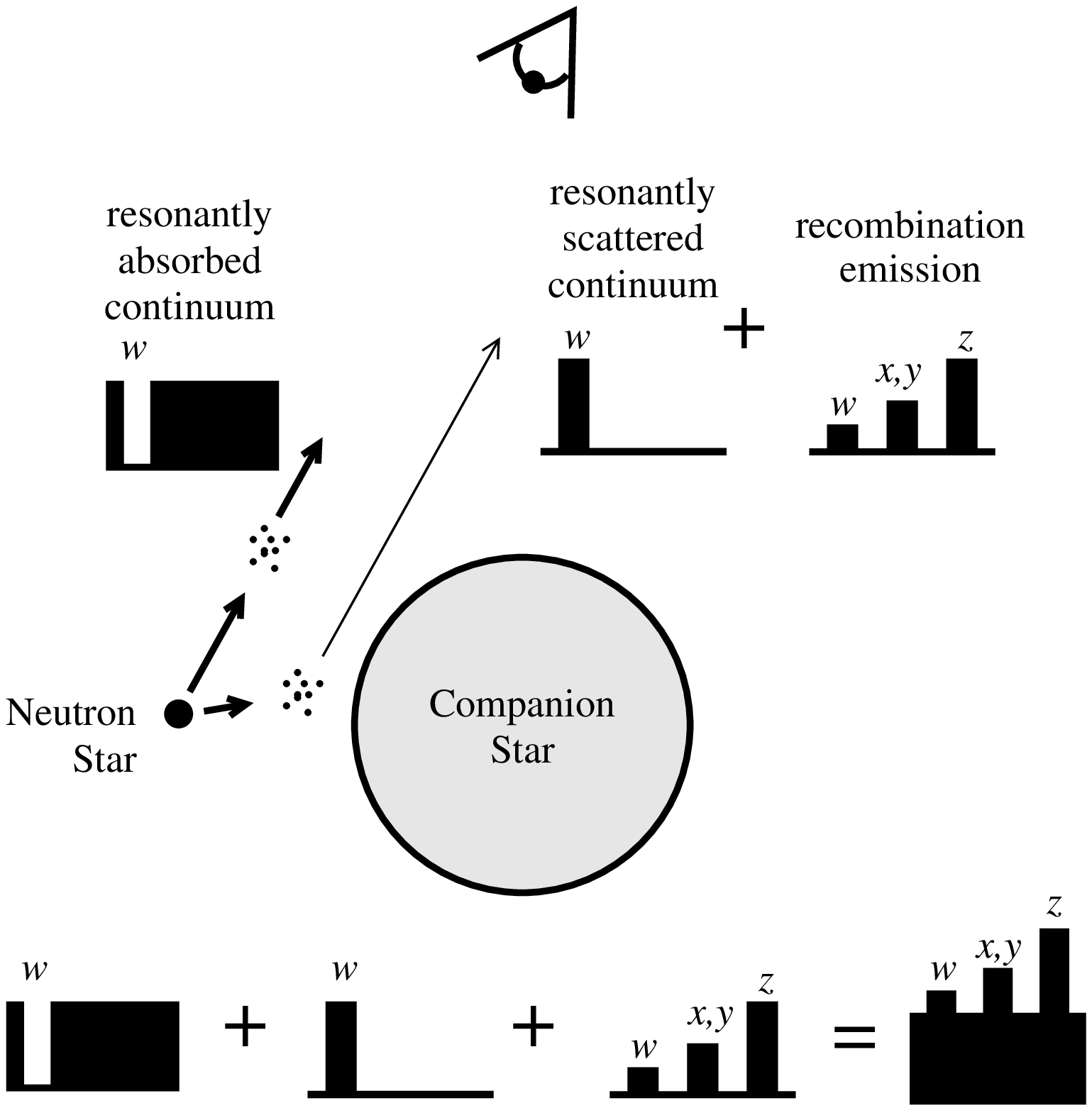} &\includegraphics[width=0.45\columnwidth,angle=0]{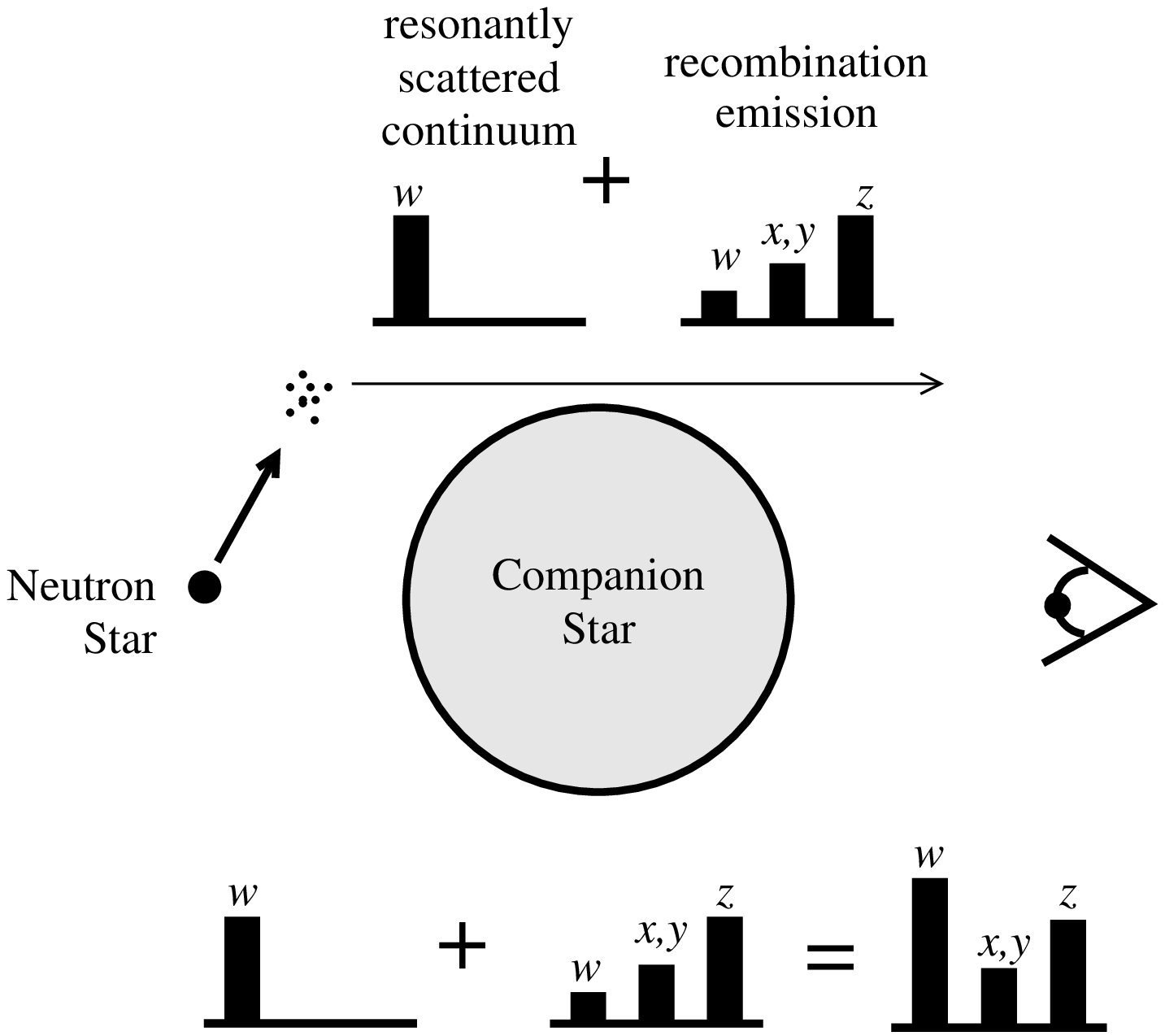}\\
\end{tabular}
\end{center}
\caption{Schemes illustrating the impact of resonance line scattering 
contribution (absorption and/or emission) on the observed \calG\ ratio
values for the high-mass X-ray binary, Centaurus X-3. 
Left and right panels show the configuration outside and during
the eclipse. Figures from \cite{Wojdowski03}. Reproduced by permission of
the AAS. }
\label{fig:RLS}
\end{figure}

At non-negligible optical depth ($\tau\sim$1), 
resonance line scattering occurs: 
a photon (from a continuum or a line) 
at a certain wavelength can be absorbed by ions, 
and then re-emitted due to a transition at the same wavelength
but generally in a different
direction. The scattered resonant photons emerge into the direction of the lowest
optical depth, then decreasing the observed value of \calG, while in 
the direction of the highest optical depth, the observed value of
\calG\ is raised. Therefore the total photon intensity integrated
over all directions remains unchanged but the photon distribution with respect to a given
direction is altered. Thus its effects depends on the geometry of the
region being observed. One should notice that in case of a spherical
source resonant line scattering does not affect the emergent spectrum. 
Resonance line scattering first occurs in transition with large
oscillator strengths, such as the so-called resonance lines. 
From \ion{C}{v} to \ion{Si}{xiii}, in case of a
collisional plasma, the resonance line becomes sensitive 
to resonant lines scattering for  
$N_{\rm H}\sim$ 10$^{21}$--10$^{23}$\,cm$^{-2}$; 
while the intercombination and
forbidden lines are only sensitive for much higher column density: 
$N_{\rm H}\sim$ 10$^{25}$--10$^{26}$\,cm$^{-2}$, and 
$N_{\rm H}\sim$ 10$^{30}$--10$^{31}$\,cm$^{-2}$, respectively
\citep{P01}. 
Depending of the geometry of the plasma (and of the
radiation source if spatially distinct), the apparent intensity of the
resonance line can be enhanced, decreasing the observed \calG\ ratio
value. Thus a higher temperature plasma could erroneously be inferred. 
In case of a photo-ionization, a collisional plasma could be mimicked
by an apparent low value of \calG\ \citep[e.g.,][]{Schulz02}.
This is nicely illustrated by \cite{Wojdowski03},
who found that in the case of the high-mass X-ray binary, Centaurus
X-3, the \calG\ ratio outside the eclipses is consistent with the
value expected in photo-ionized plasma (or more correctly ``pure'' recombination plasma); 
while during the eclipse the \calG\ ratio is much lower and
consistent with the value expected in case of a collisional
plasma. However, they explained this apparent change of the \calG\
ratio as due to the effects of
resonance line scattering (in addition to the recombination process). 
Indeed, outside the eclipse both the absorption
and the emission component of the resonance line scattering process
are observed, then ``cancel out'' the apparent contribution of this
additional process (see fig.~\ref{fig:RLS}, left panel); while
during the eclipse only the emission 
component of the resonance line scattering is observed, increasing the intensity
of the resonance line, and then decreasing the value of the observed
\calG\ ratio (see fig.~\ref{fig:RLS}, right panel), mimicking a hot plasma. 
This is the same effect as observed in obscured active galactic nuclei where only the
emission part of the resonance line scattering is observed, then
leading in some cases to a \calG\ ratio smaller than expected for a ``pure''
photo-ionized plasma (e.g., NGC\,1068: \citealt{Kin02}, \citealt{Brinkman02}; \citealt{Guainazzi07}).
 For more details about resonance line scattering, optical depth and
column density impacts, see also 
\citet{Schmelz97} \cite{Wood00}, \cite{P01,Porquet02}, \cite{Ness03a},
\cite{Testa04b,Testa07}, \cite{Bianchi05}, \cite{Waldron07},
\cite{Leutenegger07}, \cite{Porter07}, \cite{Argiroffi09},
and Churazov et al. 2010 (this volume).

In conclusion, in case of non-negligible resonance line scattering and
optical depth, additional plasma diagnostics should be used to
discriminate between a collisional plasma and a photo-ionized one,
such as those based on the width measurement of the recombination
recombination continuum (RRC) features \citep{Liedahl96,Liedahl99}, 
and on the line ratios of Fe\,L lines \citep{Liedahl90, Kallman96}.

\newpage
\subsection{Non-ionization equilibrium plasmas}\label{sec:NIE}

\begin{figure}[!b]
\begin{center}
\begin{tabular}{cc}
\includegraphics[width=\columnwidth,angle=0]{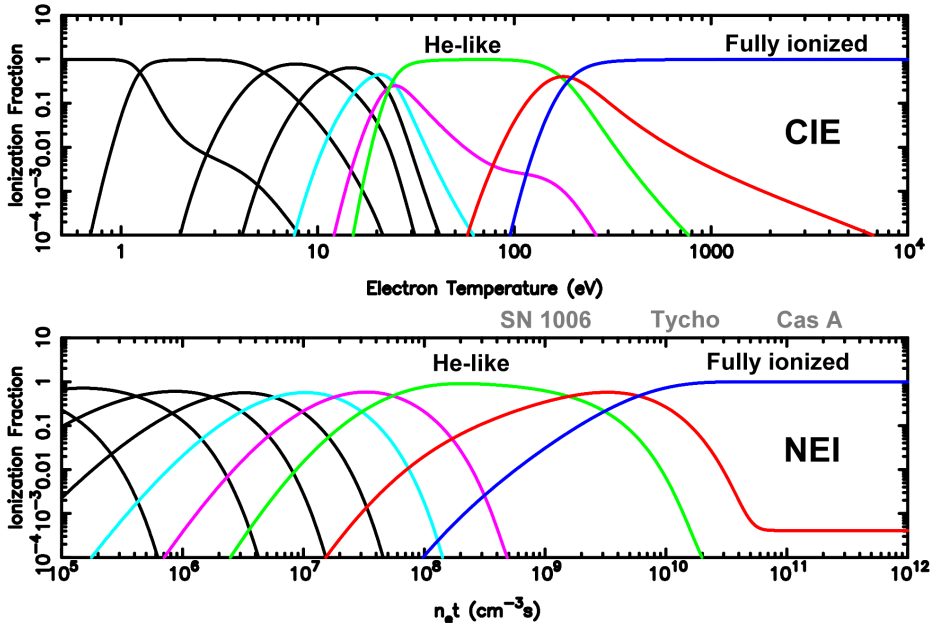} 
\end{tabular}
\end{center}
\caption{\ion{O}{vii} ionic fraction abundances for 
a collisional plasma (CIE), and a non-ionization equilibrium plasma
(NIE) for a fixed electronic temperature of 1.5\,keV. 
Figure from \cite{Vink06}. Courtesy of ESA publication.}
\vspace*{-0.5cm}
\label{fig:ionic}
\end{figure}

We now consider non ionization equilibrium (NIE) that 
occurs when the physical conditions of the source, such as the
temperature, suddenly change (transient plasmas). 
In cases where the time-scale of temperature fluctuation is shorter
than the time-scale of ionization equilibrium, the
assumptions of ionization equilibrium is not fulfilled. 
Indeed, it takes a finite time for the
plasma to respond to the temperature change, with 
different atomic process time-scales: for example at high temperature 
electron excitation time-scales are shorter than electron
ionization time-scales that are shorter than recombination time-scales.  
Since, both ionization and recombination rates depends on the
electronic 
density, time to fulfill ionization equilibrium depends on its value. 
Typically hot plasmas are out of equilibrium for 
$n_{\rm e}t \lesssim$10$^{10}$--$10^{13}$\,cm$^{-3}$\,s (e.g.,
\citealt{Smith10}).   
Therefore, the smaller the density is, the longer plasma is in
non-ionization equilibrium. 
The variations with time of the line intensities of the He-like ions
indicates the succession through the consecutive ionization stages 
and the impact of the different population level processes
\cite[e.g.,][]{Mewe75,Mewe78b,Acton78,Liedahl99,Oelgoetz04}. 
For example, an abrupt increase of the temperature can occur during
solar/stellar flares or after a shock in a supernova remnant. 
The plasma is initially ionizing, and 
the state of ionization is lower than the equilibrium
value at the local temperature. As illustrated in Fig.~\ref{fig:ionic}, the
ionic fractions calculated for a NIE plasma 
differ drastically from that found in CIE plasma.

At the early stage of the temperature increase, 
a significant abundance of Li-like ions can exist at high temperature 
together with He-like ions, contrary to that found found in CIE plasma.  
Therefore, inner-shell ionization of Li-like ions (1s$^{2}$\,2s) can be
important, especially for high-$Z$ ions, 
and then favoring the population of the $^3S_1$ level of the
He-like ions \citep[e.g.,][see also section~\ref{sec:atomic}]{Mewe75,Mewe78b,Mewe03b}.
Hence in such case, the \calG\ ratio can be higher than for a CIE
plasma of the same temperature. We would like to notice that in such
case Li-like satellites lines should be observed too, especially for
high-$Z$ ions (see section~\ref{sec:satlines}). 
Such conditions can be found at the beginning of a solar/stellar flares
\citep[see e.g.,][]{Mewe75,Mewe78b} 
or behind the shock front in supernova remnants (e.g, SN 1006: \citealt{Vink03}). 
One should notice that in case of
efficient inner-shell ionization of Li-like ions, 
the value of the ratio \calR\ is enhanced as well 
compared to that found for ionization equilibrium plasmas. \\

\begin{figure}[!t]
\begin{center}
\begin{tabular}{cc}
\includegraphics[width=0.6\columnwidth,angle=-90]{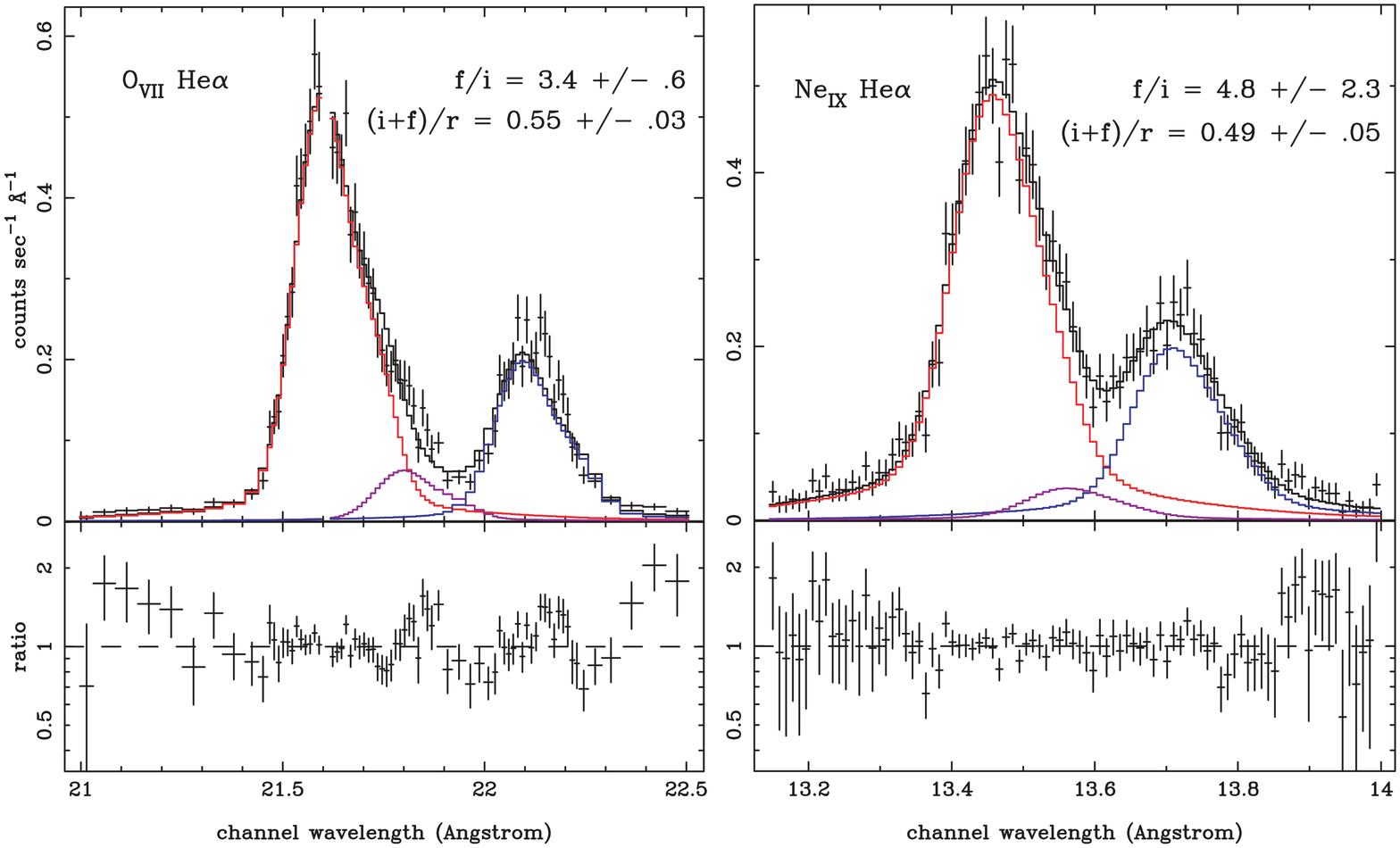} 
\end{tabular}
\end{center}
\caption{\ion{O}{vii} and \ion{Ne}{ix} lines observed from the
supernova remnants 1E 0102.2-7219 . The inferred values of \calR\
and \calG\ ratios are reported. The small values of the \calG\ ratio
indicate that the plasma is ionizing. Figures from \cite{Rasmussen01}.
Courtesy of Astronomy \& Astrophysics.}
\vspace*{-0.3cm}
\label{fig:SNR}
\end{figure}

At higher temperature when the abundance of Li-like ions becomes negligible
and the plasma is still ionizing, 
 recombination can be partly or totally quenched, especially when the
ionic fraction of H-like ions is negligible.
In such case the \calG\ ratio is
much smaller than that found for a CIE plasma at the same
temperature.
The temperatures inferred from the \calG\ ratio correspond to
temperatures well above the maximum emission temperature found for a
collisional equilibrium plasma. 
Such low \calG\ values have been found in several supernova remnants (Puppis A:
\citealt{Winkler81}; Cygnus Loop: \citealt{Vedder86}; 1E 0102.2-7219: 
\citealt{Rasmussen01,Flanagan04};
G292.0+1.8: \citealt{Vink04}). For illustration, the triplet ratios for
\ion{O}{vii} and \ion{Ne}{ix} observed for the supernova remnant 1E 0102.2-7219 are shown in
Fig.~\ref{fig:SNR}. \\

In a recombining (or cooling) plasma the ratio of H-like to
He-like ions may be much higher than that corresponding to the local
electronic temperature. Therefore, the recombination is enhanced and then would
preferentially populate the triplet levels. And depending of the
density, either the intensity of the forbidden line 
or the intercombination line(s) is
boosted compared to a CIE plasma of the same temperature. 
Such conditions may occur during a decay 
phase of the temperature during solar/stellar flares
\citep[e.g.,][]{Doschek70,Mewe78b} or during the
expansion phase of supernova remnants. \\

In conclusion, the relative intensities of the 1s\,2p-1s$^{2}$ 
He-like transitions, from \ion{C}{v} to \ion{Fe}{xxv}, 
could in principle provide a powerful test 
of departure from ionization equilibrium, 
i.e., heating/cooling stage in transient plasmas.
Complementary diagnostics can be used likewise to assess departure from 
ionization equilibrium such as 1s\,$n$p-1s\,2p ($n>$2) ratios, 
satellite lines (see section~\ref{sec:satlines}), and Fe\,L lines. 

\subsection{Charge exchange/transfer}\label{sec:CE}

X-ray lines can also be produced by transfer/exchange collisions.  
When a neutral atom/molecule ($N$; e.g., H, H2, N2, CH4, H2O, CO, CO2)
collides with a highly charged ions (X$^{+q}$; e.g., H-like ions), 
one or more electron from the neutral atom/molecule is transferred 
to the ions (eq.~\ref{eq:CE1}). The recombining ion is usually in an excited state
(X$^{+(q-1)*}$; e.g., He-like ions) and will 
be stabilized by radiative cascade (eq.~\ref{eq:CE2}).
\begin{equation}\label{eq:CE1}
X^{+q} + N \rightarrow X^{+(q-1)*} + N^{+}
\end{equation}
\begin{equation}\label{eq:CE2}
X^{+(q-1)*} \rightarrow  X^{+(q-1)} + h\nu
\end{equation}

\begin{figure}
\begin{center}
\includegraphics[width=0.8\columnwidth,angle=0]{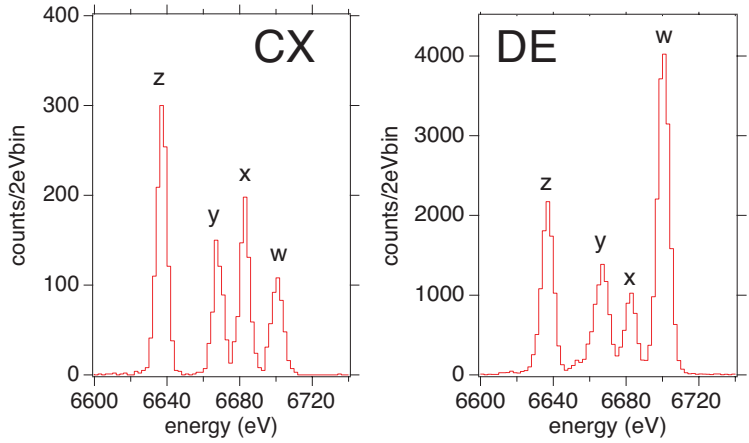}
\end{center}
\caption{Experimental spectra of \ion{Fe}{xxv} obtained with the LLNL
SuperEBIT facility for charge exchange (CX) between \ion{Fe}{xxvi}
and N$_{2}$ (left panel) and for direct electron impact excitation
(right panel). 
Figure from \cite{Brown09}. Courtesy of Journal of Physics Conference Series.}
\label{fig:CE}       
\end{figure}

This process has been proposed first proposed by \cite{Cravens97} 
to explain the unexpected extreme ultraviolet and soft X-ray emission from the 
comet Hyakutake in 1996 brought to light by ROSAT \citep{Lisse96}.
The brightest lines observed in cometary spectra are the He-like transitions of
\ion{O}{vii}, produced by the charge exchange of solar wind H-like \ion{O}{viii} ions
 with neutral targets \citep{Bhardwaj07}.  
This process has also been proposed to explain partly or totally 
the soft X-ray emission of several astrophysical objects: 
 comets, atmospheres of planets (e.g., Mars halo), Jupiter's
aurora, heliosphere, 
 terrestrial magneto-sheath, north polar spur, 
stellar winds, supernova remnants, and intra-cluster gas (e.g., \citealt{Cravens00, 
Dennerl97, Dennerl06, Kras04, Bhardwaj07, Branduardi07, Lallement04, Fujimoto07, Koutroumpa07}).  
Charge exchange cross-sections at solar wind ion energies are quite large, 
typically about 10$^{5}$ times larger than those for electron
collisional excitation. Therefore charge exchange process 
can be very efficient even for a small abundance of neutral hydrogen. 
Recent calculations and experiments show that the forbidden line
 is the brightest line of the He-like ions (e.g.,
\citealt{Beiersdorfer03, Beiersdorfer05c, Kharchenko03, Pepino04,
Wargelin05, Bodewits07,Brown09}).  
This is due to the fact that three-quarter of the captures occur into
triplet states and radiate preferentially by allowed cascade
transition into the $^{3}$P levels (that can partly -- or totally for
the $^{3}$P$_{0}$ level-- decay to the $^{3}$S$_{1}$ level) and the
$^{3}$S$_{1}$ metastable level.  
Therefore even in case of moderate spectral resolution of shift of the
energy centroid of the triplet lines towards that of the forbidden
lines (or that of the intercombination line in case of high density,
see section~\ref{sec:Te}) could be the signature of charge transfer (or
photo-ionized plasma as discussed in section~\ref{sec:CIEPIE}).  
As an illustration, Figure~\ref{fig:CE} shows the experimental spectrum of
\ion{Fe}{xxv} obtained with the LLNL SuperEBIT facility for both
charge exchange and electron-impact excitation \citep{Brown09}. 

In case of \ion{O}{vii}, values of the \calG\ ratio have been
estimated to about 12.5, 6.7 and 4.0  by
\cite{Kharchenko03}, \cite{Pepino04}, \cite{Bodewits07}, respectively. The value of the line ratios 
depend on the nature of ions (atomic number and ionization state) 
and of the neutral target (i.e, H, He, ...) and its
velocity. Therefore high-spectral resolution coupled with high S/N spectrum
 should be able to distinguish between capture from Hydrogen or from
Helium.    

\begin{figure}
\begin{center}
\includegraphics[height=0.6\columnwidth,clip]{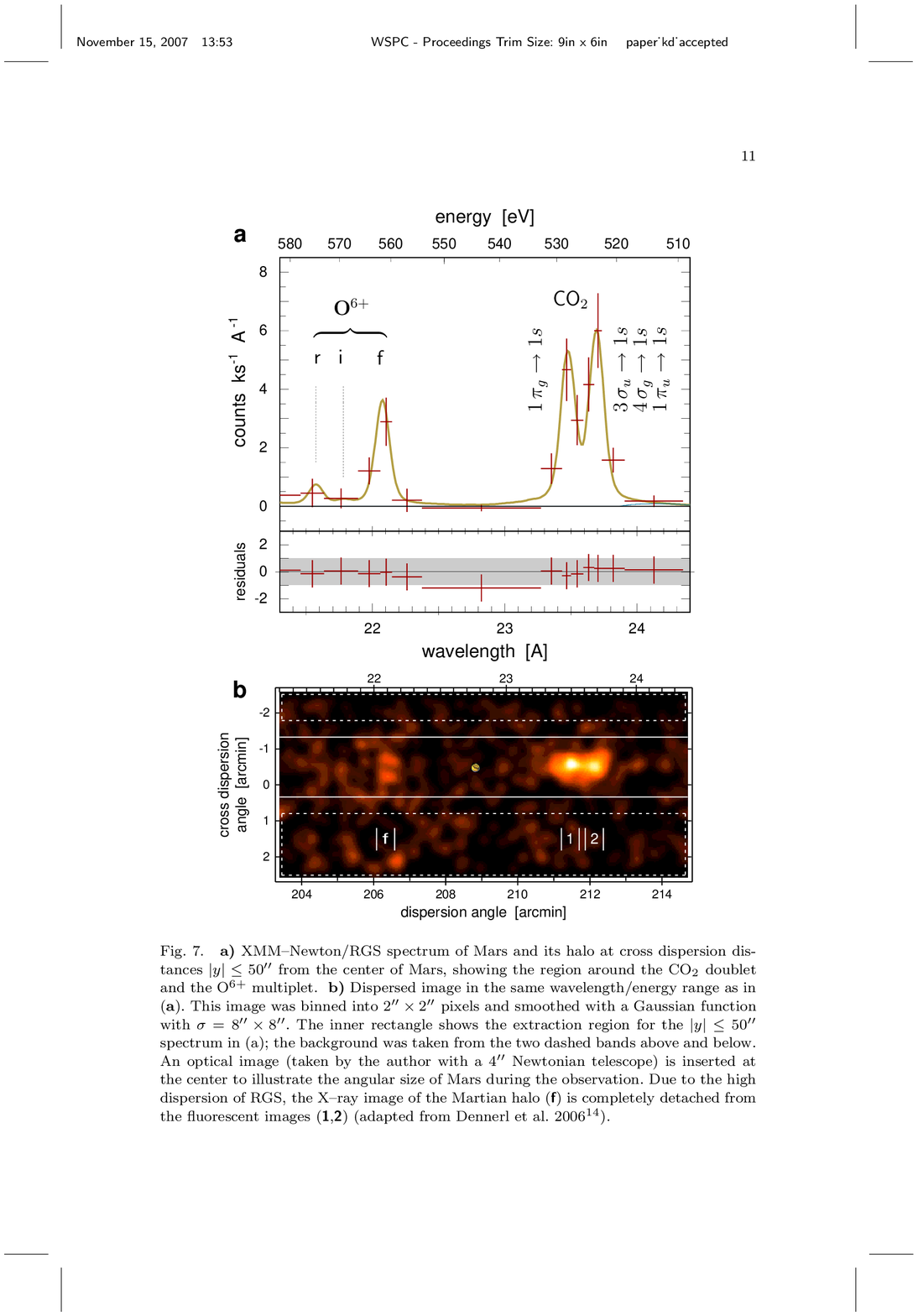}
\end{center}
\caption{XMM-Newton/RGS\,1 spectrum of Mars and its halo at cross
dispersion distances $\vert y\vert \leq$ 50\arcsec\ from the center of
Mars, showing the \ion{O}{vii} triplet \citep{Dennerl06}. Figure from
\cite{Dennerl09}. Courtesy of Advances in Geosciences.}
\label{fig:Mars}       
\end{figure}

Recently, Mars was observed with the RGS aboard XMM-Newton, 
and the \ion{O}{vii} triplet was resolved \citep{Dennerl06}. 
The spectrum of Mars halo shows the forbidden line is the dominant line of the triplet, 
with no evidence of the presence of the intercombination lines, 
and only an indication for the presence of the resonance line (see fig.~\ref{fig:Mars}). 
This is consistent with the signature of charge exchange as the source 
of the X-ray emission. The \calG\
values derived from the Mars spectra are \calG$\sim$6 for $\vert
y\vert\le 50$\arcsec.
This ratio value is consistent with the above mentioned predicted value derived for
charge exchange. However, as mentioned
by \cite{Dennerl06}, the exact value of the \calG\ ratio is highly uncertain, 
because only the forbidden line is statistically significant. 

For a detail review about the charge transfer/exchange process and its
astrophysical applications, 
see Dennerl et al.\ (2010, this volume) and references therein. 

\newpage
\section{Electronic temperature and density diagnostics}\label{sec:Tne}

\subsection{Electronic temperature diagnostic} \label{sec:Te}

\begin{figure}[!t]
\begin{tabular}{@{}c@{}c@{}}
\includegraphics[height=0.5\columnwidth,angle=90]{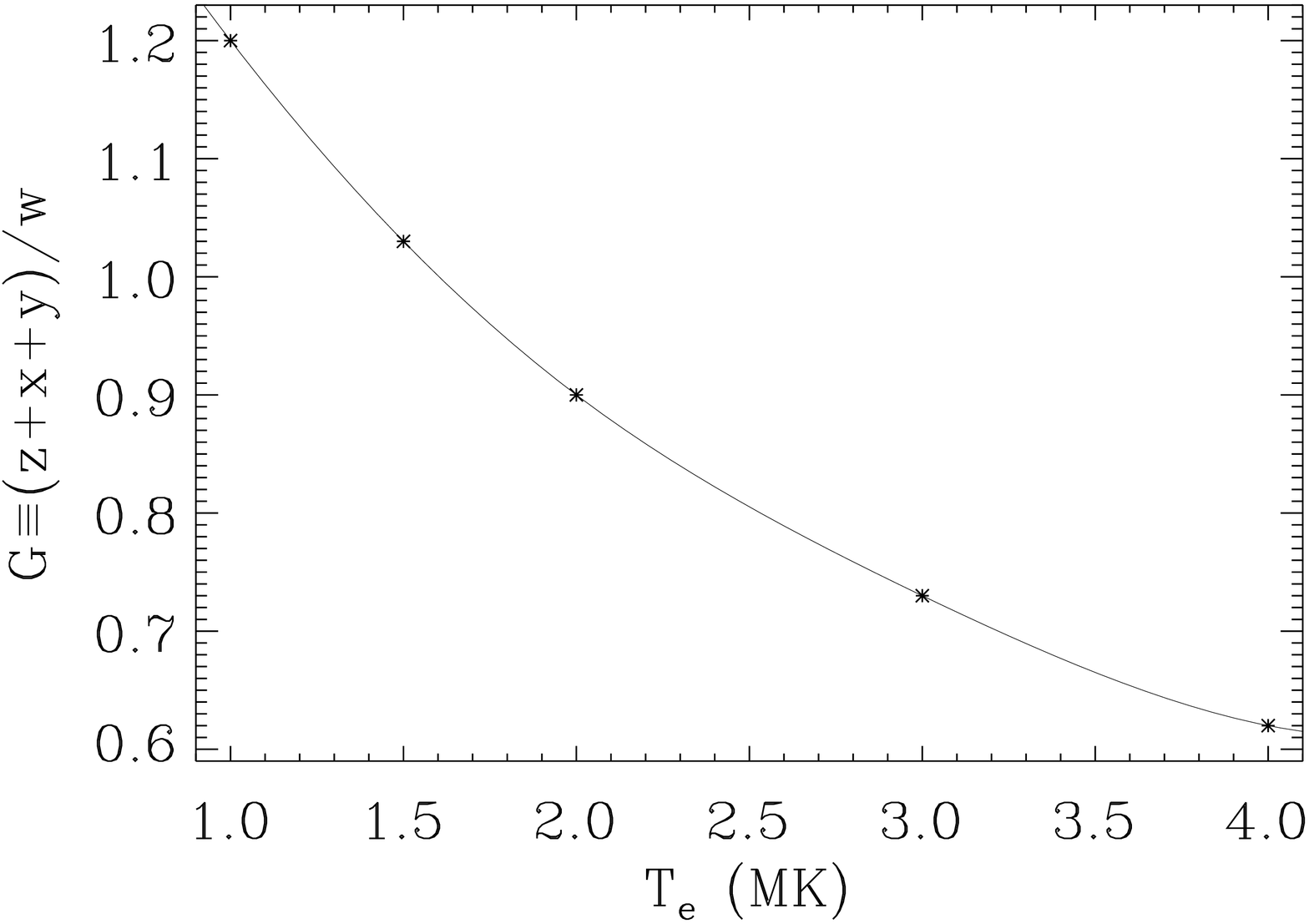}& \includegraphics[height=0.5\columnwidth,angle=90]{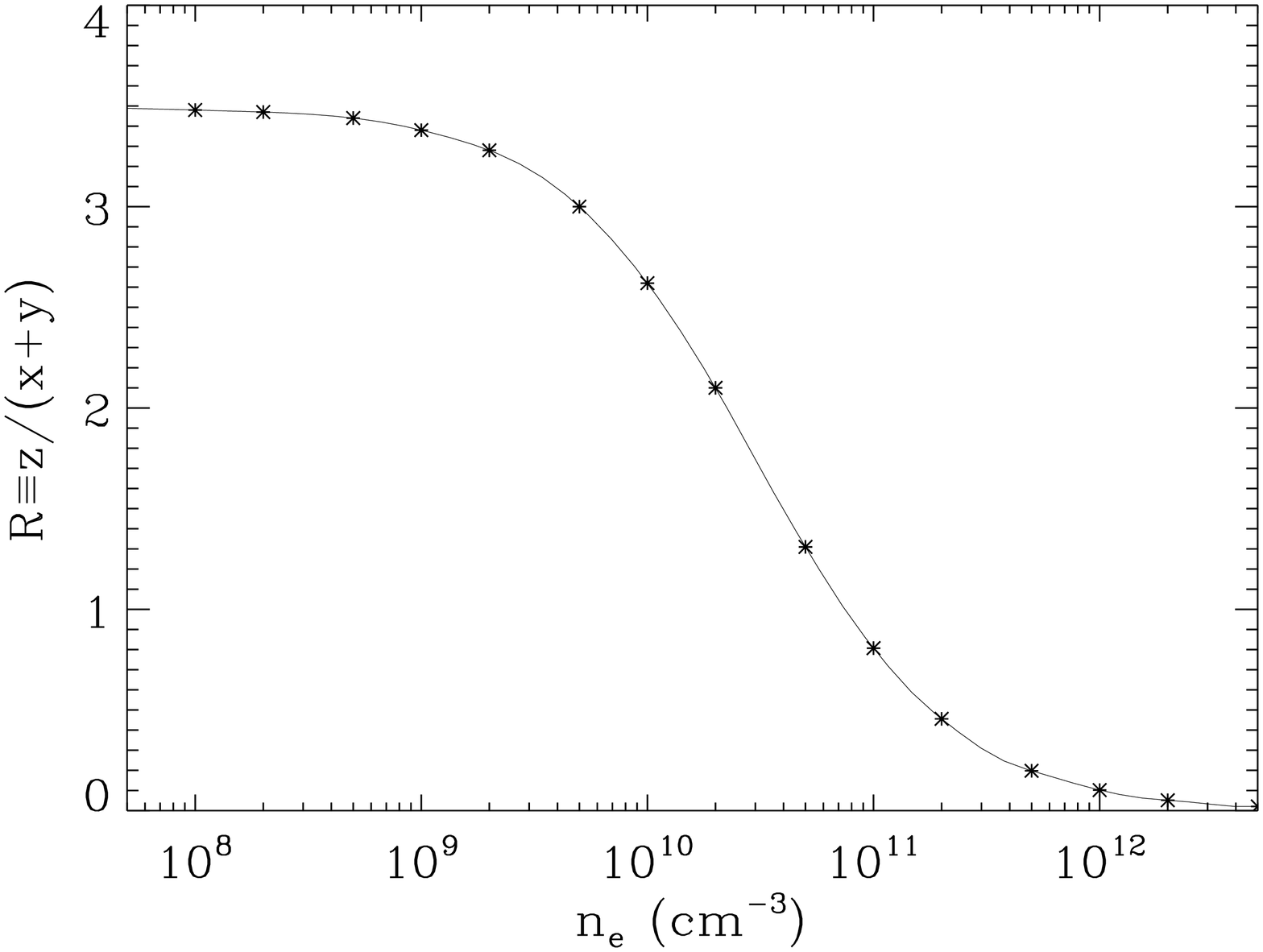}\\
\end{tabular}
\caption{Electronic temperature ($T_{\rm e}$) and density ($n_{\rm e}$) plasma diagnostics for the
He-like \ion{O}{vii} ion. {\it Left panel:} \calG\ $\equiv
(z+x+y)/w$ versus $T_{\rm e}$ for $n_{\rm
e}$=10$^{10}$\,cm$^{-3}$. {\it Right panel:}  \calR\ $\equiv z/(x+y)$ versus $n_{\rm e}$ for 
$T_{\rm e}$=10$^{6}$\,K. Calculations are taken
from \cite{P01}.}
\label{fig:OVIIGR}
\end{figure}

In the case of a collisional plasma, such as a stellar corona, the
electronic temperature can be determined using the value of the \calG\
($\equiv (z+x+y)/w$ or also $\equiv (F+I)/R$) ratio \citep[e.g.,][]{Gabriel69b,Mewe78a,PD00}. 
Indeed the collisional excitation rates do not have
the same dependence on temperature for the three lines: 
the intensity of the $w$ line increases more rapidly with the
temperature than those of the $z$ and $x$ lines (see section~\ref{sec:atomic}). 
Therefore, the \calG\ ratio decreases when the temperature
increases. As an illustration, the left panel of
figure~\ref{fig:OVIIGR} reports the \calG\ ratio versus the
temperature for \ion{O}{vii}. \\

Different temperature ranges can be probed since each ion has a specific
temperature where the line emissivity is maximum that increases with the charge of
the ion ($Z$). To illustrate this point, we report in 
Fig.~\ref{fig:ranges} the range of  electronic
temperature  where the emissivity for
the resonance line is greater than 10$\%$ 
of the maximum, computed with the CHIANTI database version 6.0.1 
\citep{Dere97,Dere09} using the CIE calculations of \cite{Bryans09}. 
When the plasma temperature increases low-$Z$
elements are easily totally ionized, so only He-like ions
from high-$Z$ elements are present. In the most favorable case
where the plasma temperature is about 2\,MK, up to five He-like ions
(from \ion{N}{vi} to \ion{Si}{xiii}) can be present simultaneously in
the plasma. 
However in case of resonance line scattering or large optical depth,
the intensity of the resonance line can be enhanced or reduced (see section~\ref{sec:CIEPIE}), 
therefore the determination of the temperature can be biased. 

Moreover, the \calG\ ratio is independent of the density when the density is
low enough to prevent:\\
$-$ the depopulation of the 1s\,2s $^1$S$_1$ level to the 1s\,2p $^1$P$_1$ level (see
section~\ref{sec:atomic});\\
$-$ the depopulation of the $n$=2 level by collisional ionization;\\
$-$ the collisional transfer from triplet to singlet term.\\

As mentioned in section~\ref{sec:atomic}, the contributions of blended
satellite lines (that depend upon spectral resolution) must be
taken into account in the calculations of the intensity of the He-like
lines \citep[e.g.,][]{P01,Sylwester08}. 
In the near future, some satellite lines will be resolved for
extra-solar objects and their intensity ratio over the resonance line 
will provide a useful temperature diagnostic (see section~\ref{sec:satlines}). 
In addition, the relative intensity of the 1s$^{2}$--1s\,$n$p (with $n
>$2) lines to the 1s$^{2}$--1s\,2p $^1$P$_1$ resonance line can be used for
temperature diagnostics \citep{Gabriel73,Keenan85}. \\

The ratio of the resonance lines of H-like/He-like ions is widely used as well for
temperature diagnostics, since this ratio increases rapidly with the
temperature. Contrary to the \calG\ ratio that is based on the
line ratio of only one ion, the H-like/He-like ratio gives 
 the temperature where both He-like ions and the more highly ionized
H-like ions are formed. When using this ratio, one assumes that the
two ions are formed in the same region and in the same volume, which
may not be the case. Moreover, this ratio depends on the accuracy of 
ion fraction calculations, and any departures from ionization
equilibrium will lead to a misleading estimate of the temperature. 
As also mentioned for the \calG\ ratio, some line blending, resonance
line scattering or optical depth effects could occur and should then
be taken into account.

\newpage
\subsection{Electronic density diagnostic}  \label{sec:ne}

\subsubsection{In the case of negligible UV radiation field}
\begin{figure}[t]
\centering
\begin{tabular}{c}
\includegraphics[width=0.5\textwidth]{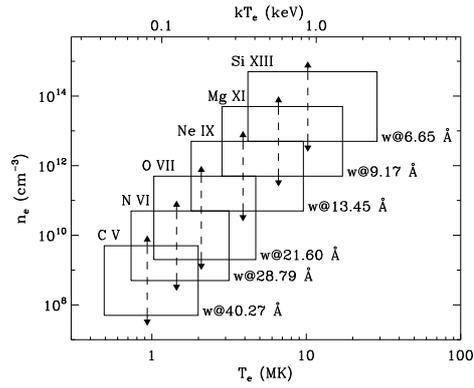} \\
\end{tabular}
\caption{Domain of electronic density and temperature where the He-like triplet diagnostics
(from \ion{C}{v} to \ion{Si}{xiii}) can be used in case of collisional
ionization equilibrium (CIE). The box 
width gives the range of electronic temperature where  the emissivity for
the resonance line is greater than 10$\%$ 
of the maximum, computed with the CHIANTI database version 6.0.1 
\citep{Dere97,Dere09} using the CIE calculations of \cite{Bryans09}. 
The dashed lines mark the electronic temperature where the maximum of the emissivity for the resonance line is achieved. The box height reports the density ranges of the ${\cal R}$ ratio diagnostic, where a constrained value of the electronic density can
be obtained. Below and above the boxes, only an upper limit and a
lower limit of the electronic 
density can be obtained, respectively. 
The wavelengths of the resonance lines are reported.}
\label{fig:ranges}
\end{figure}

The density diagnostic is based on the \calR\ ratio
($\equiv~z/(x+y)$ or $\equiv~F/I$). In the
low-density limit, both upper levels of the forbidden transition and
 of the intercombination transitions ($^3$S$_{1}$ level
and $^3$P$_{0,1,2}$ levels, respectively) decay radiatively. The
relative intensities of these lines are then independent of the
density. This is the so-called ``low density limit'' (\calR$_{0}$). 
The value of \calR$_{0}$ decreases with the atomic weight of the ions 
(e.g., \citealt{P01,Bautista00}). 
When the electronic density, $n_{\mathrm{e}}$, increases from
the low-density limit, and reaches the critical
density\footnote{The critical density $n_{\mathrm{crit}}$ is the
value of density above which collisional excitation from the 
$^3$S$_{1}$ level to the $^3$P$_{0,1,2}$ begins to dominate the
radiative decay from the metastable $^3$S$_{1}$ level to the ground
level. In some works, the critical density is defined as the density
where \calR\ is about \calR$_{0}$/2 \citep[e.g.,][]{Guedel09}.}, 
 the collisional excitation starts to
depopulate the upper level of the forbidden line to the upper levels
of the intercombination lines. Consequently, the intensity of the {\it
forbidden} line decreases while those of the {\it intercombination}
lines increase, thus implying a reduction of the ratio \calR\ over
approximately two or three decades of density.
The right panel of Fig.~\ref{fig:OVIIGR} shows the intensity 
ratio of the forbidden line over the intercombination line for 
\ion{O}{vii} versus the electronic density (see also Fig.~\ref{fig:OVIIlines} for
illustration). As explained previously, in 
the low-density regime, \calR$_{0}$ is insensitive to the density
value, and here $z\gg x+y$ (with the value of  
\calR$_{0}$ depending upon the He-like ion considered as mentioned
above; for iron \calR$_{0}$ is about unity so $z\sim x+y$). In this
case, we can infer an {\sl upper limit} for the  
density value (e.g., $\sim$ a few $10^{9}$\,cm$^{-3}$ for
\ion{O}{vii}). For $n_{\rm e}$ greater than the critical 
density, the intercombination line flux increases while the forbidden
line flux decreases. Therefore the \calR\ ratio is very sensitive to the
electronic density, and one can inferred a precise value of $n_{\rm e}$. For
example, for the \ion{O}{vii} ion reported in Fig.~\ref{fig:OVIIGR}, a
ratio equal to 2 corresponds to a density value of 
$3 \times 10^{10}$\,cm$^{-3}$. When $(x+y)\gg z$ (i.e., $z$
tends to zero), one can infer only a lower limit for the
density value, e.g., about $10^{12}$\,cm$^{-3}$ for \ion{O}{vii}. 
We report in Fig.~\ref{fig:ranges} the typical
ranges of plasma density where He-like triplets, from \ion{C}{v} to
\ion{Si}{xiii} ions, are {\it very} sensitive to the
density. These diagnostics based on 
He-like ions can probe a large density range from a few
$10^7$\,cm$^{-3}$ (\ion{C}{v}) to a few $10^{17}$\,cm$^{-3}$ (\ion{Fe}{xxv}). 
For a CIE (or PIE) plasma, high-$Z$ ions measure high density and
high temperature (or ionization parameter), while low-$Z$ ions measure low
density and low temperature (or ionization parameter) plasmas. 

We would like to notice that the value of \calR$_{0}$ is sensitive 
to the temperature \citep{Pradhan82,PD00,Smith09}. 
Indeed, the forbidden line can be enhanced compared to the 
intercombination  lines (until $n_{\rm e}$ is lower than the critical density):\\
$-$ by radiative cascade following collisional excitations to
high $n$ ($n\geq$3) levels, at high temperature (section~\ref{sec:atomic}). \\
$-$ by inner-shell ionization of Li-like ions when both the abundance
of Li-like ions and the corresponding ionization rate are high (at high
temperature). This can occur in NIE plasmas (see 
sections~\ref{sec:atomic} and \ref{sec:NIE}).\\
$-$ by the resonance contribution of auto-ionizing resonances  
to the electron scattering cross-sections at low temperature
(section~\ref{sec:atomic}).\\ 
  
As mentioned in section~\ref{sec:atomic}, the contributions of blended
satellite lines (that depend upon spectral resolution) must be
taken into account in the calculations of the intensity of the He-like
lines, and hence for \calR\ ratio calculations \citep[e.g.,][]{P01}. 

\begin{figure}[t!]
\includegraphics[width=0.9\columnwidth]{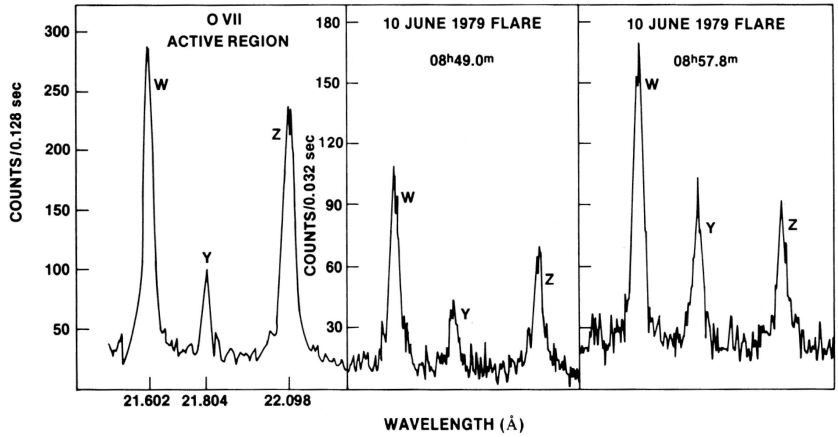} 
\caption{The \ion{O}{vii} triplet obtained
from the Sun (SOLEX spectrometer experiment on the P78-1 spacecraft)
for an active region and flares. Figure from
\cite{Doschek83}. Courtesy of Solar Physics.}
\label{fig:sun}
\end{figure}

\begin{figure}[t!]
\begin{tabular}{cc}
\includegraphics[width=\columnwidth]{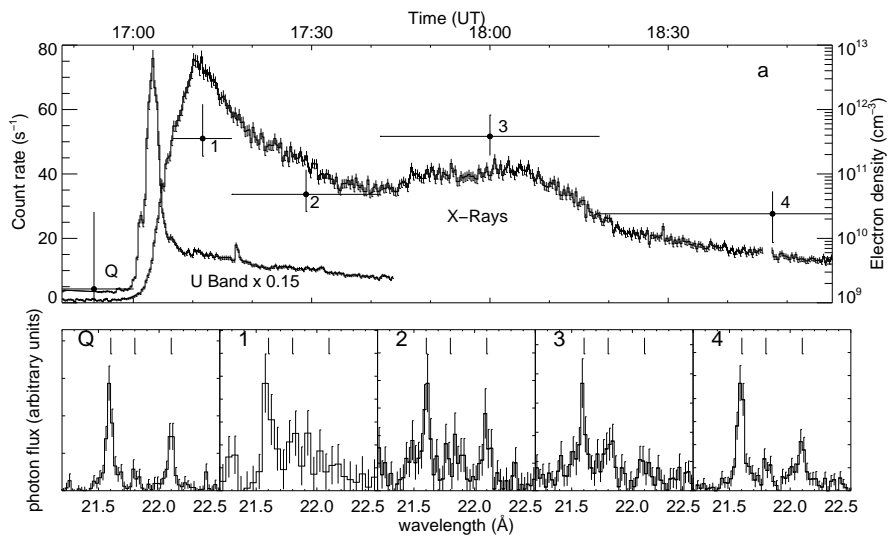}
\end{tabular}
\caption{ Evolution of the density value during a large flare on Proxima Centauri
\citep{Guedel02,Guedel04b}. The upper panel shows the X-ray and U-band light
curves. The large crosses show electronic densities inferred from
\ion{O}{vii}, during different time  
intervals (the density scale is given on the right y-axis). In the
bottom panel, the intensity of the three lines is shown for the five
time intervals. 
The three marks in the upper parts of the figures show the locations of
the resonance, intercombination, and forbidden lines, from left to
right. Figure from \cite{Guedel09}. Courtesy of  Astronomy and
Astrophysics Review.}
\label{fig:proxima}
\end{figure}

This density diagnostic has already been widely used in various type
of astrophysical objects, such as solar and stellar coronae, active
galactic nuclei, and X-ray binaries. 

In case of the Sun, the \calR\ ratio for low-$Z$ ions (e.g.,
\ion{O}{vii}, \ion{Ne}{ix}, and \ion{Mg}{xi}) show that the density for
non-flaring active regions is about 10$^{9}$\,cm$^{-3}$, while during
flares it can reach 10$^{10}$--10$^{12}$\,cm$^{-3}$
\citep[e.g.,][]{Acton72,McKenzie80,Doschek81,Pradhan81,Wolfson83,Linford88,Antonucci89}.
Figure~\ref{fig:sun} illustrates that 
the relative intensity ratio of intercombination and forbidden lines
increases during solar flares, which corresponds to a decrease of
the observed \calR\ ratio. 
This increase of the density during the flare is consistent with the 
chromospheric evaporation scenario. A similar behavior has been
observed during a large flare from Proxima Centauri
\citep{Guedel02,Guedel04b}, mainly inferred from the intensity of the
\ion{O}{vii} lines (Fig.~\ref{fig:proxima}). The intensity of the 
\ion{O}{vii}  forbidden line 
almost tends to zero during the flare,  while a strong
intercombination line stands out.  
The inferred density value rapidly increases from a pre-flare value of
$n_{\mathrm e}<10^{10}$\,cm$^{-3}$  to about 4 $\times$
10$^{11}$\,cm$^{-3}$ at flare peak, before declining to about 2
$\times$10$^{10}$\,cm$^{-3}$, then increases again during a secondary
peak, followed by a gradual decay. Such increase of the density during
flares has been suggested in several other stars (see the review by 
\citealt{Guedel09} and references therein).

This density diagnostic has been successfully applied to the  
classical T~Tauri star TW Hydrae ($\sim$ 10\,Myr) spectra, for which a
high density value, about 
10$^{13}$\,cm$^{-3}$, has been inferred by \cite{Kastner02} (see also
\citealt{Stelzer04,Raassen09b}). These
authors suggested that such a high density
plasma could be located in the shock of an accretion column. This was
the first time that X-rays were used to probe the accretion shock on a low-mass
star. Very recently, \cite{Brickhouse10} have reported the analysis of
a very deep observation of TW Hydrae with {\it Chandra/HETG} (total
time exposure $\sim$ 490\,ks). The use of the \calG\ and \calR\ ratios
has allowed them to reveal three distinct regions of the stellar
atmosphere: the stellar corona ($T_{\rm e}\sim$ 10\,MK), the accretion
shock ($T_{\rm e}\sim$ 2.5\,MK, and  $n_{\rm e}\sim$
3$\times$10$^{12}$\,cm$^{-3}$), 
and a  very large extended volume of warm
postshock plasma ($T_{\rm e}\sim$ 1.75\,MK, and $n_{\rm e}\sim$
5.7$\times$10$^{11}$\,cm$^{-3}$).       
High density values are also found in some active cool stars but
are associated with plasma with higher temperature, $T_{\rm e}\sim$
10\,MK \citep[e.g.,][]{Testa04a}, while in the case of TW Hydrae, the high
density is produced at a significantly lower temperature, 
compatible with what is expected for accretion shocks. 
Such high values of the electronic density have been found in several
other classical T~Tauri stars (see the review by \citealt{Guedel09} and references
therein). \\   

We would like to notice that since the emissivity of a plasma is
proportional to $n_{\rm e}^2$ (and the volume), any determination of
the density value is biased toward detections of the regions with
 the highest densities. Therefore, as for any other density
diagnostics, the observed \calR\ ratio reflects the density
distribution. As shown by \cite{Guedel04} in cases of stellar coronae, 
``the inferred density is not an average over the coronal volume'', 
but rather describe ``the steepness of the density distribution''. \\

In PIE plasmas, this diagnostic has also been applied to
obtain information about the density value (often upper limit) for
the warm absorber in active galactic nuclei between $10^{9}-10^{11}$\,cm$^{-3}$ 
(e.g., NGC\, 4051: \citealt{Collinge01}; NGC\,4593: \citealt{McKernan03};
NGC\,4151: \citealt{Schurch04}). For X-ray binaries, the S/N is in
general better, allowing a more accurate determination of the
line intensities, and thus on the \calR\ ratio. For example, 
\cite{Cottam01} found for the low-mass X-ray binary EXO\,0748-67,
using \ion{O}{vii} and \ion{Ne}{ix}, density values of 
2$\times10^{12}$\,cm$^{-3}$ and $>$7$\times10^{12}$\,cm$^{-3}$,
respectively (see also \citealt{Jimenez03}). 
For the magnetic cataclysmic variable AE Aqr, \cite{Itoh06} found 
a disagreement between the density value found from \ion{N}{vi} and \ion{O}{vii} of
$\sim$10$^{11}$\,cm$^{-3}$ and the geometrical scale (inferred from
the emission measure), and concluded that the
plasma cannot be produced by mass accretion into the white dwarf.

\subsubsection{Influence of the UV radiation field}\label{sec:UV}

\begin{table}[!b]
\begin{center}
\begin{tabular}{ccccccc}
\hline
\hline
     \ion{C}{v}  & \ion{N}{vi} & \ion{O}{vii} & \ion{Ne}{ix} &
\ion{Mg}{xi} & \ion{Si}{xiii}\\
\hline
2280  & 1906  & 1637 & 1270 & 1033 & 864 \\
\hline
\hline
\end{tabular}
\end{center}
\caption{Wavelengths in Angstrom for the UV transition between $^{3}$S$_{1}$ and
$^{3}$P$_{1}$ levels that correspond to the upper levels of the
forbidden line $z$ and the intercombination line $y$ (the 
dominant intercombination line for ions lighter than \ion{S}{xv})
 from \ion{C}{v} to \ion{Si}{xiii} \citep{P01}. }
\label{tab:UV}
\end{table}

As shown previously, the $^{3}$S$_{1}$ metastable ($m$) level
can be depopulated to the $^{3}$P level ($p_k$) by collisions with
electrons (but also in some cases by protons and
$\alpha$-particles), then increasing the relative intensity of
intercombination line over that of the forbidden line. However,
another atomic process can be responsible of a strong intercombination
line: this is the excitation due to a UV photon
 (e.g.,
\citealt{Gabriel69b,Doschek70,Blumenthal72,Mewe75,Mewe78a,P01,Kahn01,Waldron01}). 
Indeed, the wavelengths corresponding to the $^{3}$S$_{1}$ to $^{3}$P
 transitions are in UV (see table~\ref{tab:UV}). 
Therefore, a strong UV radiation field can mimic a high-density plasma. 
The rate (in s$^{-1}$) of absorption of the UV photon is:
\begin{equation}B_{mp_k} = {{W A_{p_km} (w_{p_k}/w_m)}\over {{\rm exp}\Bigl({{\Delta E_{mp_k}}\over {kT_{\mathrm rad}}}\Bigr) - 1}},
\end{equation}
where $A_{p_km}$ is the spontaneous radiative decay of the
$p_k$ level ($^{3}P_{0,1,2}$) to the $m$ level ($^{3}S_{1}$), $w_{p_k}$
and $w_m$ are the statistical weights of the $p_k$ and $m$ levels,
respectively, $\Delta E_{mp_k}$ is the energy between the $p_k$ and $m$
levels, and $T_{\mathrm rad}$ is the effective radiation
temperature, i.e., the black body temperature of the radiation source, 
and $W$ is the dilution factor of the radiation, defined as: 
\begin{equation}\label{eq:W}
W=\frac{1}{2}~\left[1-\left(1-\left(\frac{r_{*}}{r}\right)^2\right)^{1/2}\right],
\end{equation}
where $r$ is the distance from the center of the stellar source of
radius $r_{*}$. For stars such as Capella or Procyon, $W$ is equal to 0.5,
because UV from the photosphere irradiates coronal structures that are
close to the stellar 
surface. In a binary star such as Algol the UV radiation field
originates from the hot primary star that illuminates the cool
secondary star, which emits the X-rays. Therefore, in this latter case $W$ is much lower
 ($\simeq 0.01$). The impact of the UV radiation field is then dependent of the
$T_{\mathrm rad}$ value and/or the dilution factor of the radiation.\\

\begin{figure}[t!]
\begin{tabular}{cc}
\includegraphics[width=4cm,angle=90]{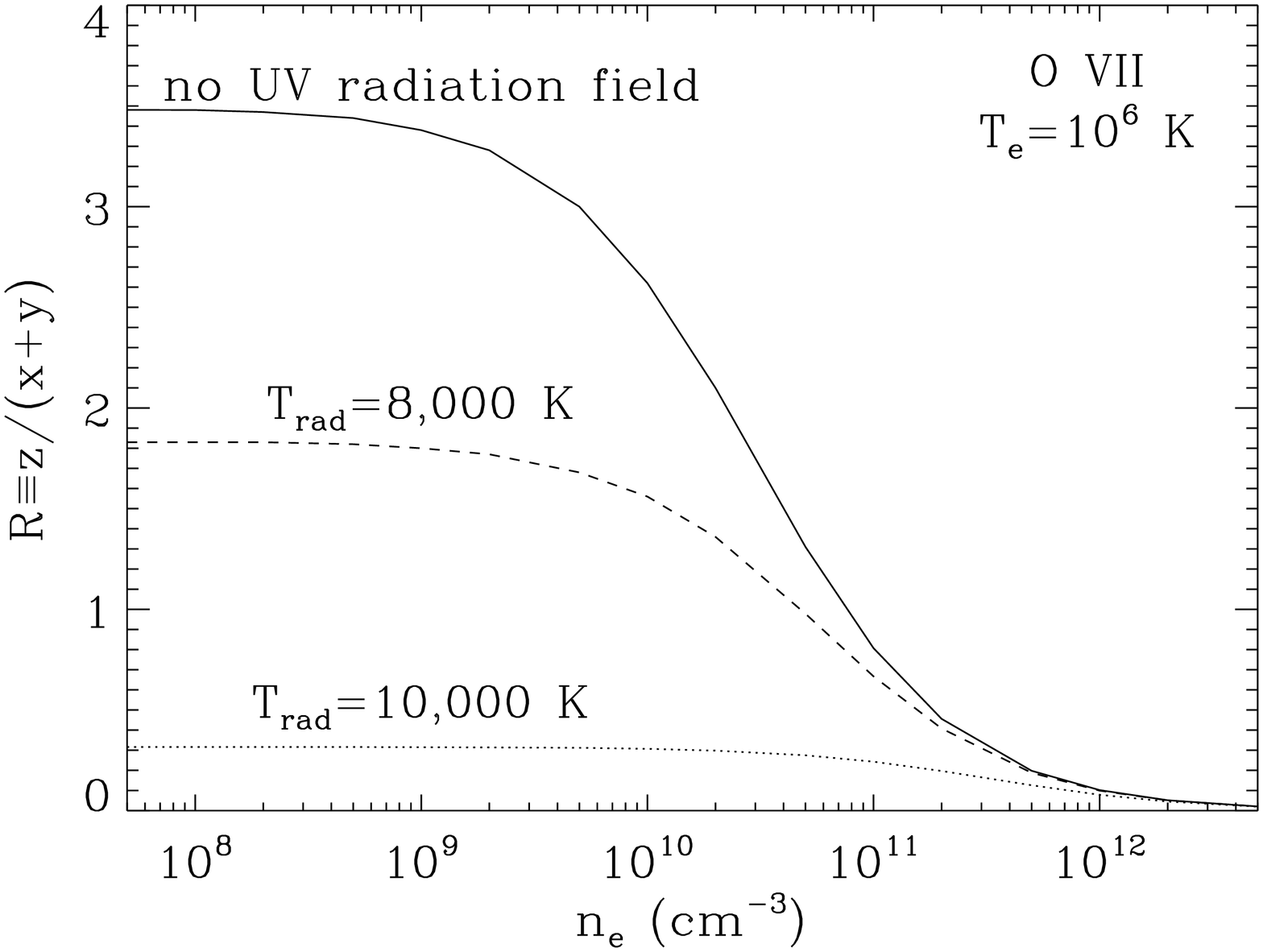} & \includegraphics[width=4cm,angle=90]{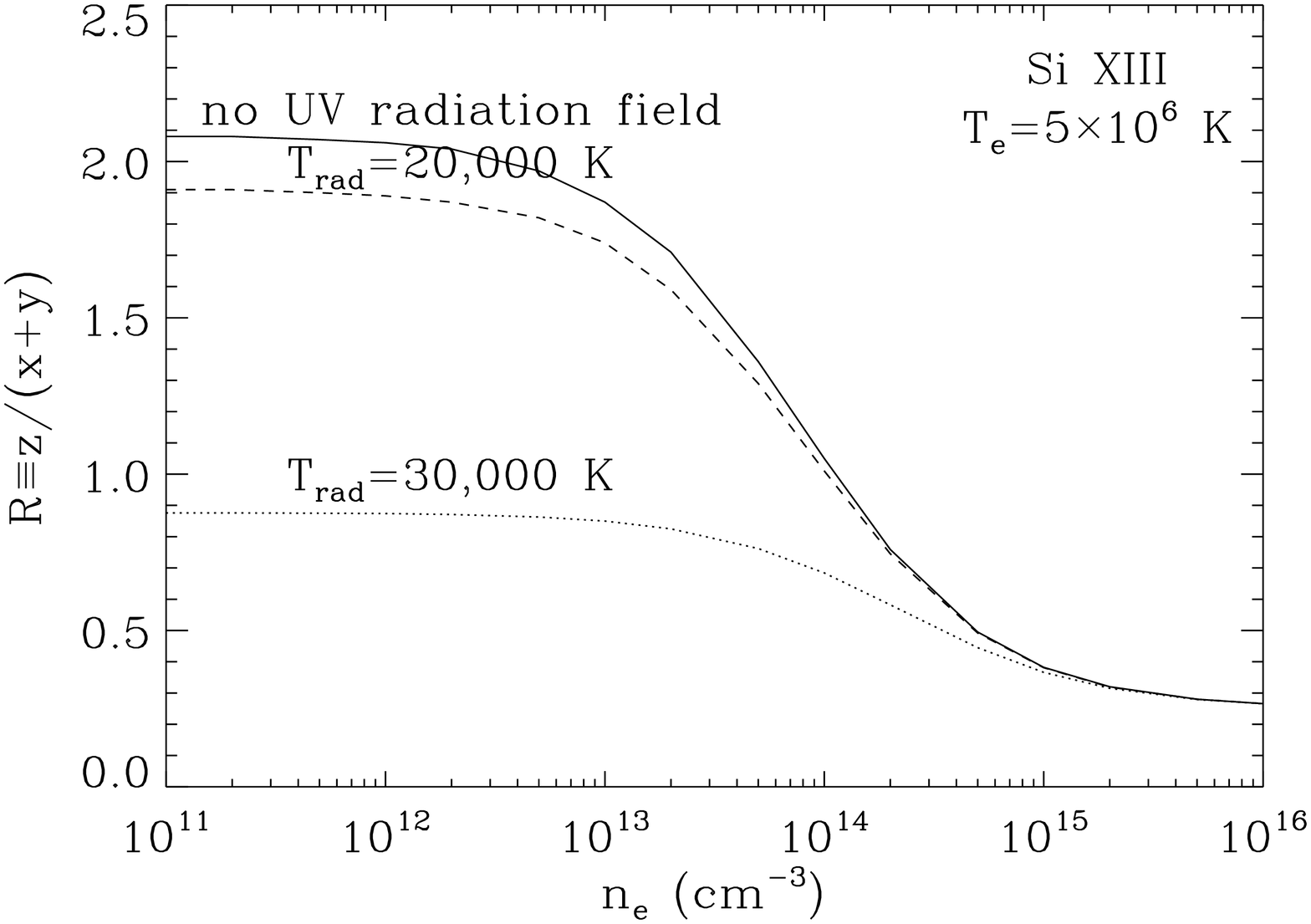}\\
\end{tabular}
\caption{Effects of UV photo-excitation on the calculated \calR\ ratios for
\ion{O}{vii} and \ion{Si}{xiii}, with a dilution factor $W$ of
0.5. The \calR\ ratio in case of no radiation field is given for
comparison.   
{\it Left panel:} \ion{O}{vii} at $T_{\rm e}$=10$^{6}$\,K and
for two values of $T_{\rm rad}$ (8\,000\,K, and 10\,000\,K).
{\it Right panel:} \ion{Si}{xiii} at $T_{\rm e}$=5$\times$10$^{6}$\,K and
for two values of $T_{\rm rad}$ (20\,000\,K, and 30\,000\,K). 
Calculations are from \cite{P01}.}
\label{fig:UV}
\end{figure}

\begin{figure}[t!]
\centering
\begin{tabular}{cc}
\includegraphics[width=5cm,angle=-90]{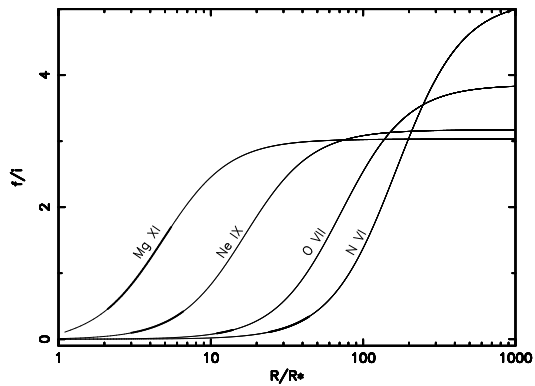}\\
\end{tabular}
\caption{The \calR\ ratio values (taking into account the UV
photo-excitation) versus the radial distance between the plasma where
the He-like ions are formed and the UV photosphere. The present
calculations, for \ion{N}{vi}, \ion{O}{vii}, \ion{Ne}{ix} and
\ion{Mg}{xi}, are performed for the O star $\zeta$ Orionis assuming $T_{\rm
rad}\sim$ 30\,000\,K. The thickened portion of each
curve shows radial distance of the X-ray plasma inferred from the
measured \calR\  ratio values. Figure from \cite{Raassen08}. 
Courtesy of Astronomy \& Astrophysics.}
\label{fig:fiversusR*}
\end{figure}

The left panel of Fig.~\ref{fig:UV} displays the \calR\ ratio for
\ion{O}{vii} (with $W$=0.5) 
when there is no radiation field and for two values of
$T_{\mathrm rad}$, i.e., 8\,000\,K and 10\,000\,K. 
The ratio decreases when the radiation field
temperature increases, especially \calR$_{0}$. 
For example, an observed line ratio of about 2
corresponds to a density of a few 10$^{10}$\,cm$^{-3}$ when the UV
radiation field can be neglected, 
and to only an upper limit about 10$^{9}$\,cm$^{-3}$
for $T_{\rm rad}$=8\,000\,K. 
The right panel of Fig.~\ref{fig:UV} shows
that the \calR\ ratio for \ion{Si}{xiii} is sensitive to higher values of
$T_{\mathrm rad}$. 

To summarize, the photo-excitation process has an important impact for low-$Z$
ions (e.g., \ion{C}{v}, \ion{N}{vi}, \ion{O}{vii}) even for moderate
$T_{\mathrm rad}$ values \citep[e.g.,][]{Mewe78a,P01}, while for
higher $Z$ ions
(e.g., \ion{Mg}{xi}, \ion{S}{xv}), this process becomes only important for
high radiation temperature ($\sim$ few 10\,000 K), as found, for example,
in hot (O and B) stars \citep[e.g.,][]{P01,Kahn01,Waldron01}. 
Indeed, \cite{Kahn01} have found that, for the O star, $\zeta$
Puppis, the $forbidden$ to $intercombination$ line ratios within the
He-like triplets are abnormally low for \ion{N}{vi}, \ion{O}{vii},
\ion{Ne}{ix}, and even for \ion{Mg}{xi}. These authors show that this
is not due to a high electronic density, but rather due to the intense
radiation field of this star. 

At the low-density limit or when the value of the density is known, 
the \calR\ ratio can then be  used to determine the distance between
the plasma where the He-like ion is found and the UV source, e.g.,
the radial distance from the X-ray source where the He-like ions are
formed and the UV photosphere in case of stars
\citep[e.g.,][]{Kahn01,Waldron01,Miller02,Mewe03a,Behar04,Henley05,
Schulz06,Leutenegger06,Raassen08}.
Figure~\ref{fig:fiversusR*} shows the \calR\ 
ratio values (taking into account the UV
photo-excitation) versus this radial distance. The present
calculations, for \ion{N}{vi}, \ion{O}{vii}, \ion{Ne}{ix} and
\ion{Mg}{xi}, are performed for the O star $\zeta$ Orionis assuming $T_{\rm
rad}\sim$ 30\,000\,K \citep{Raassen08}.
According to the UV wavelengths reported in Table~\ref{tab:UV}, 
the flux at the wavelengths corresponding to ions up to \ion{Mg}{xi}
can be determined observationally since these are longward of the Lyman
limit; while the wavelength corresponding for higher $Z$ ions  are 
shortward of the Lyman limit and thus must 
be inferred from stellar atmospheric model spectra.

For massive stars, O and early-B stars, the X-ray emission is commonly
explained as produced from shocks formed by instabilities within a radiatively
driven wind. He-like diagnostics can be used to discriminate two different possible
geometries: (1) the spatially distributed wind shock model, where the
plasma density is low and far from the UV emitting photosphere; and (2)
the magnetically confined wind shock  model \citep{Babel97},
where the plasma density is high and close to the UV emitting
photosphere. For example, the brightest stars 
of the Orion nebula's trapezium  ($\Theta^{1}$ Ori A, C and E) shows
He-like triplet emission 
consistent with the magnetic confinement model \citep{Schulz03,Gagne05}. 

This ratio has been also used to infer the UV flux emitted by the
accretion disk in the case of X-ray binaries, once the location and
density are known \citep[e.g.,][]{Jimenez03}.

\section{Satellite lines to He-like ions}\label{sec:satlines}

\begin{figure}[b!]
\centering
\begin{tabular}{cc}
\includegraphics[width=0.9\columnwidth]{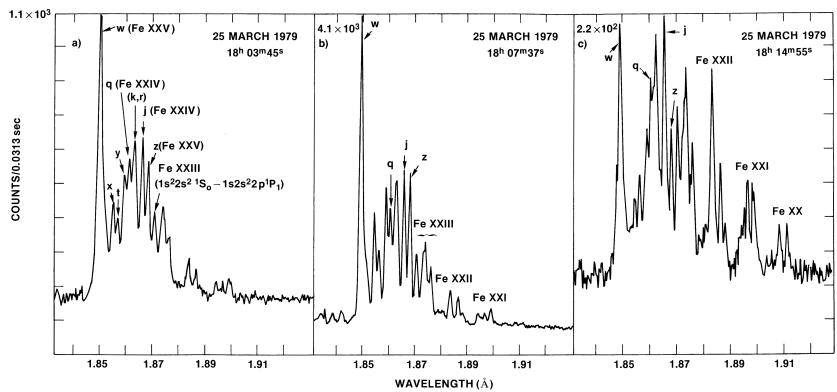}\\
\end{tabular}
\caption{Solar flare spectra of \ion{Fe}{xxv} showing the presence of
satellite lines from \ion{Fe}{xxiv},
\ion{Fe}{xxiii}, and \ion{Fe}{xxii}. Figure from \cite{Doschek79}. 
Courtesy of Astrophysical Journal.} 
\label{fig:satlinessolar}
\end{figure}

Satellite lines are lines that appear very
close to or even blended with the lines of He-like and H-like
ions (or other ions) and are formed by dielectronic recombination (DR) or inner-shell
excitations (IE) by electrons or photons. 
Satellite lines have been observed in high-resolution
solar spectra, as illustrated in Fig.~\ref{fig:satlinessolar}. 
The intensity ratios of the satellite lines to resonance line can
provide valuable diagnostics for the determination of electronic 
temperature, ionic fraction of Li-like/He-like (as well as
Be-Like/He-like), and departures from Maxwellian energy distribution \citep[see
e.g.,][]{Gabriel69a,Gabriel72,Gabriel79,Bely-Dubau79,Bely-Dubau82,Dubau80,Lemen84,Mewe99,Oelgoetz01}.
Such diagnostics will provide new diagnostic tools to probe extra-solar plasmas thanks
to the gratings and calorimeters on the {\sl Astro-H} and
{\sl IXO} satellites.  

In the first laboratory observations of He-like ion resonance 
lines of carbon by \cite{Edlen39}, some ``satellite'' lines appeared close to them. 
During the following years, Edlen and collaborators \citep[see][]{Edlen47} obtained more 
He-like spectra for fluorine, magnesium and aluminium, which showed 
that the relative intensities of these satellite lines to the resonance line increased with 
nuclear charge $Z$. They rapidly identified them as representing the same electron transition, 
1s$^2$ -- 1s\,2p, as the main line with the addition of one or possibly 2 outer electrons, 
e.g. 1s$^2$\,(2s) -- 1s\,2p(2s).
From the former identification, it was evident that some of these lines could be explained 
by inner-shell electron excitation of Li-like and Be-like ions by free electron collision. But it
was only clear after \cite{Burgess64}'s famous paper  that the other lines
were excited by DR. The first classification of satellite lines, 
according to LS terms, was made by \cite{Gabriel69a}.
But for highly ionized elements, where the fine-structure is often resolved, a new classification
was proposed by \cite{Gabriel72}, which is still used. This 
classification goes from $a$ to $v$ for the (Li-like, Be-like, ...) 
$n=2$ satellite lines  (1s$^2$\,2l' -- 1s\,2l\,2l') of ``parent'' He-like
lines;  $w$, $x$, $y$ and $z$ correspond to the parent lines, i.e.,
the resonance, intercombination and forbidden lines. Note that capital
letters are used to define He-like satellite lines to H-like parent
lines, e.g., $J$ (see below). 

\begin{figure}[t!]
\centering
\begin{tabular}{cc}
\includegraphics[width=7cm]{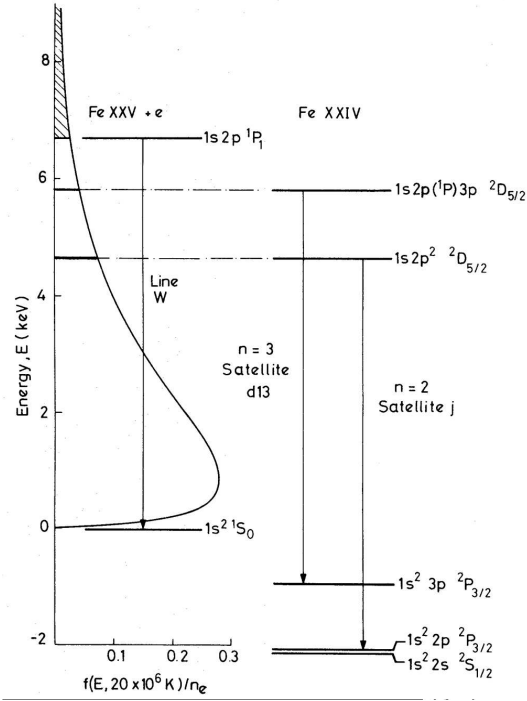}\\
\end{tabular}
\caption{Energy level diagram for the \ion{Fe}{xxv} resonance line, $w$,
and two dielectronic recombination satellite lines $w$ and $d_{13}$. 
Figure from \cite{Gabriel79}. Courtesy of MNRAS.}
\label{fig:satlines}
\end{figure}

\cite{Gabriel69a} proposed to use intensity ratios of some satellite lines 
to deduce different plasma parameters such as the electronic temperature from the
intensity ratio of a dielectronic line, e.g. $j$ and $k$ in Gabriel's notation,  to the parent resonance
line $w$. The $j$ and $k$ are the strongest $n$=2 satellite lines. 
As an example, we shall explain the dielectronic recombination process responsible
for the $j$ line intensity, i.e., 1s$^2$\,2p $^2$P$_{3/2}$ -- 1s\,2p$^2$ $^2$D$_{5/2}$. The upper level
of the satellite line lies well above the first ionization limit of Li-like elements, i.e., 1s$^2$ and
under the first excited He-like levels belonging to 1s\,2s and 1s\,2p configurations.
Due to electron-electron repulsion, it is unstable and can decay by auto-ionization :
an electron is ejected. From the energy conservation rule, the ejected electron energy $\epsilon$ can be 
deduced by $\epsilon= E({\rm 1s\,2p}^2 \,^2{\rm D}_{5/2}) -E({\rm
1s}^2 \,^1{\rm S}_0) $.   
In fact, from the quantum mechanics energy uncertainty relation, the possible energies $\epsilon$ are spread, 
with the energy width being proportional to the inverse of auto-ionizing level lifetime which 
contains both radiative and auto-ionization decays.
And, from the micro-reversibility principle, the auto-ionization process has a
reverse process called dielectronic capture (or radiationless
capture). A free plasma electron with an energy 
inside the small $\epsilon$ width, already described, can be captured
by a He-like ion in its ground level 
1s$^2$. The resulting auto-ionizing level subsequently decays either by
auto-ionization or by radiation (if possible).  
The latter possibility corresponds to the $j$ satellite line emission,
usually a ``pure'' dielectronic recombination line, i.e., only DR can
produced the $j$ line.  
The electrons contributing to the excitation of the parent line $w$ and to the satellite line $j$ have
very distinct energy ranges. The line ratio $j/w$ is therefore strongly temperature dependent. It is 
the basic principle of the diagnostic proposed in 1969. Later on,
\cite{Bely-Dubau79} calculations for $n\geq$ 3 for iron He-like satellite lines, 1s\,2l\,$n$l -- 1s$^2n$l, showed that these DR satellite lines were 
blended with $w$, and therefore contributed to the observed $w$ intensity. 
To correct the electronic temperature diagnostic, one has to suppress 
from the observed $w$  the contribution  
of unresolved dielectronic satellites, to get a meaningful line
ratio. This is particularly important  
for low electronic temperature diagnostics in CIE plasmas, i.e., much
lower than the temperature of maximum formation. 
Thus, the resolution or not of the satellite lines to the main He-like line ions
depends on the ions ($Z$), the spectral resolution and the $n$ levels 
where the lines are formed for DR, hence their contributions must be accounted
for accordingly
\citep[e.g.,][]{Gabriel79,Bely-Dubau79,P01,Oelgoetz01,Phillips08,Sylwester08}. 
Interestingly, though using data for which the spectral resolution was
insufficient to resolve DR satellite lines, 
\cite{Audard01} were the first ones to apply
temperature diagnostics based on their (lack of) contribution   
to set an lower limit to the electronic temperature of the cool plasma
of a stellar coronae, Capella. 

Since the upper level of satellite lines are auto-ionizing levels, most of them are populated by
dielectronic recombination and it is often the dominant population
process. Nevertheless, theoretical 
atomic calculations have shown that some auto-ionizing levels can have
very small auto-ionization probabilities. This is the case for the $q$
line, 1s$^2$\,2s\,$^2$S$_{1/2}$\,--\,1s\,2p($^1$P)2s $^2$P$_{3/2}$,
for example. Due to
interaction effects between electrostatic interaction and spin-orbit,
the matrix elements cancel and the auto-ionization of the $q$ line
is almost zero for $Z$ nuclear charge close to iron ($Z=26$). The $q$
line is then excited only by electron collision from the Li-like ground
state 1s$^2$\,2s (IE). The free electron energy range for
excitation being almost the same and the atomic collisional process
being similar, the ratio of $q$ over $w$, corrected for unresolved
satellites, gives directly the abundance ratio of Li-like to He-like
ions, almost independently of the electronic temperature. It is a very
reliable diagnostic that was used, conjointly with the $j/w$ or  
$k/w$ ratios, to construct  observational curves of relative
abundances of Li-like and Be-like abundances to 
He-like, versus electronic temperature, for iron and calcium from solar
data \citep[e.g.,][]{Antonucci87}.
Both electronic temperature and ionic abundance diagnostics 
were very popular in solar X-ray astronomy
to study solar flares, from 1969. An 
 exhaustive review of the solar observations made until 1979 can be
found in \cite{Dubau80}. Since 1980, one of the most
original uses of these diagnostics has been in tokamaks, 
to determine the electronic temperature in the plasma core where
the nuclear fusion might take place. 

\begin{figure}[t!]
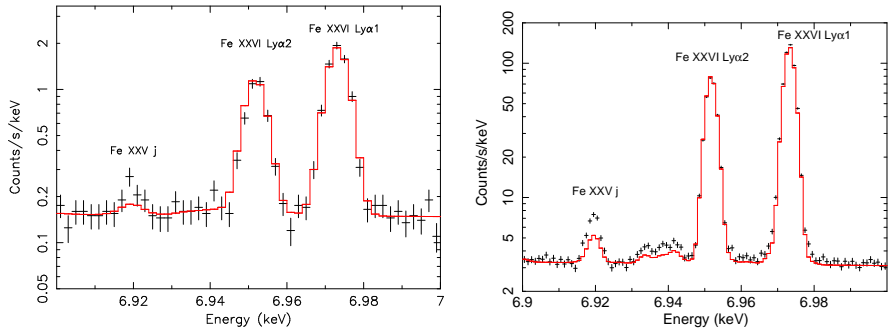

\begin{tabular}{cc}
\includegraphics[width=0.35\columnwidth,angle=-90]{Kaastr09_f4b.ps} & \includegraphics[width=0.35\columnwidth,angle=-90]{Kaastr09_f4a.ps}\\
\end{tabular}
\caption{{\sl Astro-H} (left panel) and {\sl IXO 100\,ks} (right
panel) simulations of the
spectrum expected for a supra-thermal
electron distribution behind the shock in a cluster of galaxies.
The excess detection of the $J$ satellite line of the He-like
\ion{Fe}{xxv} ion to the H-like 
\ion{Fe}{xxvi} parent resonance lines will be significantly detected. Figures from
\cite{Kaastra09}. Courtesy of Astronomy \& Astrophysics.}
\label{fig:J}
\end{figure}

Non-thermal electron distribution can have an impact
on the atomic process rates, such as electron excitation, ionization and
recombination (hence on the ionization 
balance): e.g., \cite{Owocki83}, \cite{Dzifcakova92},
\cite{Porquet01b}, \cite{Kaastra09}, \cite{Prokhorov09}.
The ratios of the satellite to
resonance line are sensitive to departure from a Maxwellian electron
distribution \citep[e.g.,][]{Gabriel79,Seely87,Mewe99}. 
Indeed the satellite lines and the resonance lines are not sensitive
to the same part of the electron distribution.
 As shown for \ion{Fe}{xxv} in Fig.~\ref{fig:satlines}, the
resonance line is produced by excitation by electrons with energy above
 6.701\,keV, while the DR satellite lines are due to the capture of
electron with an energy, to within the auto-ionization width ($>
10^{-4}$\,keV), of 4.694\,keV and 5.815\,keV for the $j$ and d13
satellite lines, respectively. Therefore, depending on the shape of
the electron distribution, either the intensity of the 
satellite lines or of the resonance
line will be modified compared to that expected from a Maxwellian electron
distribution. In case of a Maxwellian electron distribution, 
the temperature inferred from satellite lines must be identical. 
Such diagnostics will be useful to test the departure from a Maxwellian electron
distribution in , e.g., stellar flares, supernova remnants, and clusters of
galaxies. 
As an illustration, Fig.~\ref{fig:J} shows {\sl Astro-H} and {\sl
IXO} 100\,ks simulations of the spectrum expected for a supra-thermal
electron distribution behind the shock in a cluster of galaxies,
showing that the excess detection of the $J$ He-like \ion{Fe}{xxv} 
satellite line to the resonance line of the H-like \ion{Fe}{xxvi}, 
will be significantly detected \citep{Kaastra09} if fit assuming an
electron Maxwellian distribution. The so-called $J$ He-like satellite 
line is composed of several satellite lines with transitions
1s\,2$l$--2$l$\,2p $l$=s, p \citep{Dubau81}. This satellite line has
already been observed in solar flare spectra \citep{Parmar81,Tanaka86,Pike96}. 

In high temperature CIE plasmas, DR satellite
lines are usually observed in emission for high $Z$ and inner-shell
satellite lines are even present for low $Z$, (e.g., Carbon
spectra from \citealt{Edlen39}).
Conversely, in the low temperature plasmas found in PIE plasmas, both
dielectronic recombinations rates become negligible, hence  
only satellite lines formed by inner-shell photo-excitation 
 can be significant, e.g.: the 
satellite lines formed by inner-shell photo-excitation
of Li-like ions: $q$, $r$, $s$, and $t$
(1s$^2$\,2s + photon $\rightarrow$ 
1s\,2s\,2p) and of Be-like ions: $\beta$ (1s$^2$\,2s$^2$ + photon $\rightarrow$
1s\,2s$^2$\,2p). The two strongest satellite lines expected are $q$ and
$\beta$, though this depends upon the relative abundance of Li-like and
Be-like ions, respectively. Such satellite lines have already been
observed in the spectra of the Seyfert type I galaxy, NGC\,5548
\citep{Steenbrugge03}, though in absorption since in type I active
galactic nuclei the so-called 
warm absorber/emitter that are
thought to emit highly-ionized lines are seen mainly in absorption. 
In Figure 3 of \cite{Steenbrugge03},  one can see, close to the oxygen
He-like lines (\ion{O}{vii}), inner-shell satellite lines of Li-like (\ion{O}{vi}) and
Be-like (\ion{O}{v}), in absorption. One must also mention that the parent
\ion{O}{vii} $w$ line is also observed (partly) in absorption. 
In Seyfert 2 galaxies, the satellite lines (as for the main He-like lines) should be
observed in emission, allowing the use of the above diagnostics.\\

Very high-spectral resolution spectra up to an energy of 10\,keV that
will be obtained  in the 
near future will allow observers to apply these diagnostics for numerous
astrophysical plasmas such as stellar coronae, supernova remnants, X-ray
binaries, active galactic nuclei, and clusters of galaxies.

\section{Conclusions and perspectives}\label{sec:conc}

As reviewed here, diagnostics based on the main He-like ion lines can be
very useful to determine the properties of astrophysical plasmas, 
such as the ionization processes (collisional excitation,
recombination, charge transfer), departure from ionization
equilibrium, electronic temperature,
 electronic density, and distance between the plasma
where the He-like ions are formed and the UV radiation source (e.g.,
for stellar photosphere). The main advantages of
the use of these close lines is that, for a given He-like ions, 
the line ratios are emitted in the same emitting volume, 
and are moreover insensitive to instrumental calibration, 
 Galactic column density effects, and elemental abundances. These diagnostics were
first applied to the solar corona for both active regions and flares.
Thanks to the current generation of X-ray satellites, {\sl Chandra} and
{\sl XMM-Newton}, these diagnostics are successfully used to probe the
physical properties of extra-solar plasmas such
as stellar coronae (from cool to hot stars), active galactic nuclei, 
X-ray binaries, and supernova remnants. \\

With the next generation of X-ray satellites, namely {\sl Astro-H} and
{\sl IXO}, unprecedented spectral resolution combined with higher
sensitivity will be reached. For {\sl Astro-H} (JAXA, launch planned in
2014) with the Soft X-ray Spectrometer (SXS) aboard, 
a FWHM spectral resolution of at least
7\,eV (with a goal to 4\,eV) over the 0.3--12\,keV energy range will
be attained. While for {\sl IXO} (ESA, NASA, JAXA), the planned 
resolving power of the
grating (XGS) will be at least 3\,000 for the soft 0.3--1\,keV
energy band, and a spectral resolution of 2.5\,eV 
for the X-ray Microcalorimeter Spectrometer (XMS)
 over the broad 0.3--7\,keV energy range. 
Therefore the resonance, intercombination and forbidden lines of
He-like ions will be resolved up to \ion{Ni}{xxvii}, including
\ion{Fe}{xxv}.  
Therefore higher density and/or higher temperature plasmas will be
probed such as, for example, in magnetic cataclysmic variables, X-ray
binaries, active galactic nuclei, and galaxy clusters. 
The resolution of several satellite lines will permit
 useful diagnostics for the determination of electronic
temperature, ionic fraction, departure from ionization equilibrium and
 from Maxwellian electron distribution. 
Moreover, the combination of higher spectral resolution and sensitivity will be
used to determine various physical properties of numerous astrophysical
objects, including weak and high-redshift ones. Then statistics of the
physical properties of a certain classes of objects (active galactic nuclei, X-ray
binaries, galaxy clusters, stellar corona, supernova remnants) will
be performed over 
luminosities, types, accretion rates, and distances. \\

We would like to conclude this review with the following advice from
\cite{Liedahl99}: ``In general, we need to bring to bear as many
diagnostics as are available in order to make an internally consistent
interpretation of an X-ray spectrum''.

\begin{acknowledgements}
The authors would like to sincerely thank the anonymous referee for
his thorough reading of this manuscript. 
D.~P.\ would like to acknowledge the organizing committee 
for inviting her to give this review about He-like ions, 
and would like to dedicate this review to 
Rolf Mewe.
\end{acknowledgements}

\bibliographystyle{aps-nameyear}      
\bibliography{dporquet}   
\nocite{*}

\end{document}